\documentclass[aps,prd,secnumarabic,preprint,showpacs]{revtex4-1} 

\usepackage{mathrsfs,amsmath,amssymb,bm,braket,here}

\usepackage{color, graphicx}
\usepackage[colorlinks=true]{hyperref}

\begin{document}

\preprint{KOBE-COSMO-17-02}
\title{Photon-Axion Conversion, Magnetic Field Configuration, \\ and  Polarization of Photons}%

\author{Emi Masaki}%
\email[]{emi.masaki@stu.kobe-u.ac.jp}

\author{Arata Aoki}%
\email[]{arata.aoki@stu.kobe-u.ac.jp}

\author{Jiro Soda}%
\email[]{jiro@phys.sci.kobe-u.ac.jp}

\affiliation{Department of Physics, Kobe University, Kobe 657-8501, Japan}%
\date{\today}


\begin{abstract}
We study the evolution of photon polarization during the photon-axion conversion process with focusing on the magnetic field configuration dependence.
Most previous studies have been carried out in a conventional model where a network of magnetic domains is considered and each domain has a constant magnetic field.
We investigate a more general model where a network of domains is still assumed, but each domain has a helical magnetic field.
We find that the asymptotic behavior does not depend on the configuration of magnetic fields.
Remarkably, we analytically obtain the asymptotic values of the variance of polarization  in the conventional model.
When the helicity is small, we show that there appears the damped oscillating behavior in the early stage of evolution. 
Moreover, we see that the constraints on the axion coupling and the cosmological magnetic fields using polarization observations are affected by 
 the magnetic field configuration. 
 This is because the different transient behavior of polarization dynamics is caused by
 the different magnetic field configuration.
Recently, [C. Wang and D. Lai, J. Cosmol. Astropart. Phys. 06 (2016) 006.] claimed that the photon-axion conversion in helical model behaves peculiarly.
However, our helical model gives much closer predictions to the conventional discontinuous magnetic field configuration model.

\end{abstract}

\maketitle
\tableofcontents

\section{Introduction}

It is widely known that the dark matter problem is an important issue in cosmology.
A psuedo-scalar field, the so-called axion, has been a candidate of the dark matter.
The axion was originally introduced to resolve the strong CP problem of QCD (QCD-axion)~\cite{QCD1,QCD2,QCD3,QCD4}.
In this case, the mass of the axion is related to the axion coupling constant.
Theories beyond the standard model of particle physics, such as string theory, also predict pseudo-scalar fields (axion-like-particles) for which there is no relation between the mass of axion and the axion coupling constant~\cite{String1,String2}.
In this paper, we do not distinguish them, and we call general pseudo-scalar fields axions.
There are interesting cosmological and astrophysical phenomena associated with the axion which can be used to constrain the axion mass and the coupling constant.

Another important issue in cosmology is the primordial magnetic fields.
In fact, there are several evidences for the existence of magnetic fields on cosmological scales, {\it i.e.}, intergalactic magnetic fields (IGMF).
The Faraday rotation measurements and observations of the cosmic microwave background radiations give the upper limit $|\bm{B}_{IGMF}|\lesssim 10^{-9}\rm{G}$~\cite{Magnetic_Field1,Magnetic_Field2}.
Moreover, the measurements of $\gamma$-ray give the lower limit $|\bm{B}_{IGMF}|\gtrsim 10^{-16}\rm{G}$~\cite{experiment3}.
Theoretically, it is not straightforward to explain the primordial magnetic fields with the cosmological coherent length.
Although several mechanisms including the generation of magnetic fields during inflation are suggested, it is fair to say there exists no convincing mechanism for producing cosmological magnetic fields.

It is well known that the axions can be converted to photons and {\it vise versa} in the presence of magnetic fields~\cite{a-p:1,a-p:2}.
Therefore, the photon-axion conversion might shed light on the above two issues.
Indeed, cosmological consequences of this mechanism have been widely investigated~\cite{Ia:1,Ia:2,Ia:3,Ia:4,Ia:5,transparency:1,transparency:2,transparency:3,transparency:4,transparency:5,transparency:6,transparency:7,transparency:8,transparency:9,transparency:10,others:1,others:2,others:3,others:4,helical:1,helical:2,helical:3}.
It was suggested that the dimming of supernovae can be explained by mixing with axion~\cite{Ia:1,Ia:2,Ia:3,Ia:4,Ia:5}.
Mixing can also increase the transparency of high energy photon~\cite{transparency:1,transparency:2,transparency:3,transparency:4,transparency:5,transparency:6,transparency:7,transparency:8,transparency:9,transparency:10}.
In addition, the astrophysical and cosmological consequences have been studied from various points of view~\cite{others:1,others:2,others:3,others:4}.  Conventionally, one assumes that the universe is made by patches of coherent domains, each with a uniform magnetic field and that their direction varies randomly from domain to domain.
However, it is possible for magnetic fields to have more general configurations.
Indeed, if the cosmological magnetic fields are generated during inflation in the presence of parity violating terms, the resultant magnetic field can be helical~\cite{Magnetic_Field1,Magnetic_Field2}.
Hence, it is worth studying the photon-axion conversion process in more general magnetic field configurations.
In this paper, we study the magnetic field configuration dependence of the photon-axion conversion process with a model where a network of magnetic domains are assumed, each has a helical magnetic field, and its helicity randomly changes from domain to domain. 
Recently, Wang and Lai~\cite{helical:3} claimed that the photon-axion conversion in helical model behaves peculiarly.
However, as we will see in section III.2, we find that our helical model gives much closer predictions to the conventional discontinuous magnetic field configuration model.
Moreover, we give the detailed analysis of the evolution of polarization of photons.

We also consider implication of our results for the astronomical and cosmological phenomena.
As a demonstration, we compare our theoretical analysis with the polarization measurements of gamma-ray bursts (GRBs) to constrain model parameters.
As is well known, GRBs, the brightest events in the universe, occur a few times per a day.
After the $\gamma$-ray emission, they have also broadband long-lasting radiation in the X-ray, optical, and radio wavelengths ranges, which is called afterglow.
The measurement of polarization covers the energy range from about 10\,keV to 100\,keV.
The standard model of afterglows is the synchrotron emission which can produce the linear polarization of photons.
Indeed, all the observations show that the circular polarization of GRBs is less than 1\%~\cite{experiment4}.
Since the photon-axion conversion in magnetic fields can produce the circular polarization, we can constrain the strength of magnetic fields so as not to exceed the observational limit.
In this paper, we discuss the configuration dependence of constraints on the axion coupling constant and the strength of IGMF.

The paper is organized as follows.
In section \ref{2}, we review the conversion mechanism by assuming a single domain.
Then, we discuss conversion probability in the conventional model where a constant magnetic field is assumed in each domain of a network of domains.
Interestingly, we succeed in analytically deriving asymptotic values of the variance of  polarization, which show a good agreement with numerical results.
In section \ref{3}, we investigate evolution of photon polarization in the photon-axion conversion process with a helical magnetic field in each domain.
As a demonstration, we give constraints on the model parameters.
We also discuss the effects of the magnetic field configuration.
The final section \ref{conclusion} is devoted to the conclusion.

\section{Photon-Axion Conversion in Conventional Model}\label{2}

In this section, we provide an overview of photon-axion conversion.
We calculate the conversion rate in a uniform magnetic field.
Next, we consider the conventional model, namely, many coherent domains, each with a uniform magnetic field, but the field changes randomly from domain to domain.
We explain how to obtain asymptotic averaged values of intensity of the axion and the photon.
We show averaged polarization vanishes, and obtain the variance of polarization analytically.

\begin{figure}[h]
\begin{center}
\includegraphics[width=5cm]{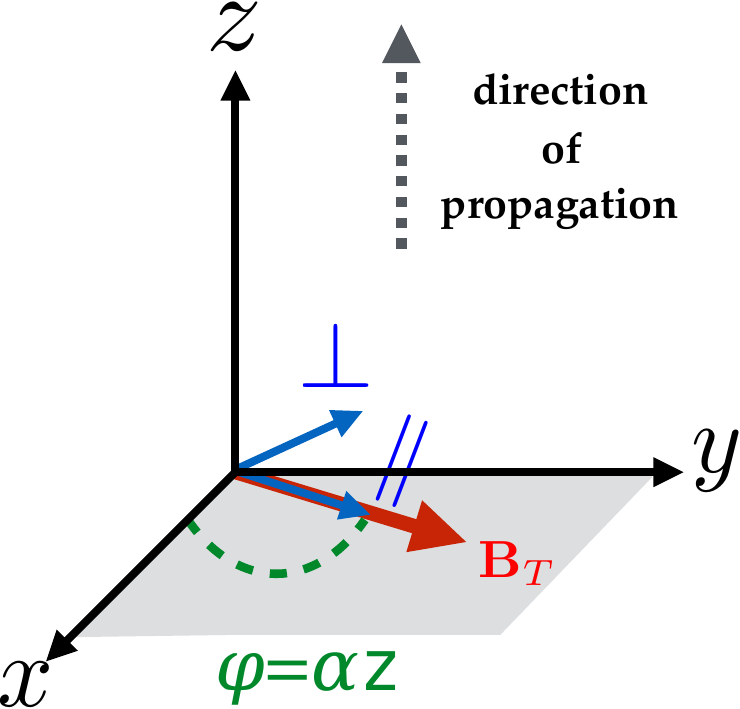}
\caption{Coordinate System}
\label{fig1}
\end{center}
\end{figure}

We consider the following photon-axion system
\begin{equation}\label{action}
  S\ =\ \int d^4 x
  \left[ - \frac{1}{2}(\partial _\mu a\, \partial^{\mu}a + m_a^2 a^2) \\
  -\frac{1}{4}F_{\mu \nu} F^{\mu \nu} -\frac{1}{4}\,g_{a\gamma\gamma}\,aF_{\mu \nu}\tilde{F}^{\mu \nu} \right], 
\end{equation}
where $a$ is an axion field with mass $m_a$, $F_{\mu \nu} \equiv \partial_{\mu}A_{\nu} - \partial_{\nu}A_{\mu}$ is the field strength of the electromagnetic field $A_{\mu}$, $\tilde{F}_{\mu \nu} \equiv \frac{1}{2} \varepsilon_{\mu \nu \rho \sigma}F^{\rho \sigma}$ is its dual, and $g_{a\gamma\gamma}$ is a coupling constant with the dimension of inverse energy.
One can divide electromagnetic field into constant external magnetic field $\bm{B}$ and the propagating photon field $A_\mu = (0,\bm{A})$.
Hereafter we choose radiation gauge $\nabla \cdot \bm{A}=0$, and we write $\bm{A}$ and $a$ as plane waves
\begin{equation}
  \bm{A}(z,t)\ =\ i \left(
     \begin{array}{c}
       A_1(z)\\
       A_2 (z)\\ 
       0 \\    
     \end{array}
    \right) e^{- i \omega t}\ ,\quad 
    a(z,t)= a(z)\,e^{- i \omega t},
\end{equation}
provided that magnetic fields vary in space on scales much larger than photon or axion wavelength.
As is illustrated in Fig.~\ref{fig1}, $(1,2)$ denotes $(x,y)\,\text{or}\,(\parallel,\perp)$.
We consider a monochromatic light beam traveling along the $z$-direction.
$\bm{B}_T$ denotes the projection of $\bm{B}$ in the $x$-$y$ plane and $A_\parallel$, $A_\perp$ denote components of the photon field parallel and perpendicular to $\bm{B}_T$ respectively.

Let us show how the photon and the axion convert each other.
The photon-axion conversion is sensitive to the magnetic field structure.
However, we assume that the direction of external magnetic field is constant, and its strength is homogeneous.
In fact, the photon-axion conversion in the constant magnetic field is a good starting point.

From the action (\ref{action}), we can derive the Klein-Gordon equation
\begin{equation}
     (\Box -m_a^2)\,a\ =\ \frac{1}{4}\,g_{a\gamma\gamma} F_{\mu \nu} \tilde{F}^{\mu \nu}\ ,
\end{equation}
and Maxwell's equation
\begin{equation}
    \partial_\mu F^{\mu \nu}\ =\ -g_{a\gamma\gamma}\,\tilde{F}^{\rho \nu}\partial_\rho a \ .
\end{equation}
One can divide these source terms of equations into background magnetic field $\bm{B}_T$ and the propagating photon $\bm{A}$.
The dispersion relation is assumed to be $\omega\simeq k$.
We can obtain linearized equations of motion
\begin{equation}
  \begin{cases}
    \left[(\omega+i\partial_z) - \dfrac{m_a^2}{2\omega}\right]a\ =\ -\dfrac{1}{2}g_{a\gamma\gamma}\, B_T A_\parallel\ , \\
    (\omega + i \partial_z) A_\parallel\ =\ -\dfrac{1}{2}g_{a\gamma\gamma}\,B_T\,a\ , \\
    (\omega + i \partial_z) A_\perp\ =\ 0\ ,
   \end{cases}
\end{equation}
where we used $\Box\,=\,(\omega+i\partial_z)(\omega-i\partial_z)\,=\,(\omega+i\partial_z)(\omega+k)\,\simeq\,2\omega(\omega+i\partial_z)$.
Now, we introduce $\Psi$
\begin{equation*}
  \Psi\ \equiv\ \left(
     \begin{array}{c}
       a (z)\\
       A_\parallel(z)\\
       A_\perp(z)
     \end{array}
   \right) e^{-i\omega z } \ .
\end{equation*}
Then, we obtain Schr\"odinger-like equation
\begin{align}
  \begin{split}
    &i\frac{d}{d z}
    \Psi
  =       
     \left(
      \begin{array}{ccc}
        \dfrac{m_a^2}{2\omega} & -\dfrac{1}{2}g_{a\gamma\gamma}B_T & 0 \\
        -\dfrac{1}{2}g_{a\gamma\gamma}B_T& 0& 0 \\
        0&0&0
      \end{array}
     \right)
     \Psi\ .
   \end{split}
\end{align}

In reality, we need to take into account the Euler-Heisenberg effective Lagrangian which gives the first quantum correction to the Maxwell equations
\begin{equation}
  \mathcal{L}_{EH}\ =\ \frac{1}{90m^4_e}\left(\frac{e^2}{4\pi}\right)^2
  \left[\left(F_{\mu\nu}F^{\mu\nu}\right)^2+\frac{7}{4}\left(\tilde{F}_{\mu\nu}F^{\mu\nu}\right)^2 \right]\ ,
\end{equation}
where $m_e$ is the electron mass, $e$ is the electron charge and we choose natural Lorentz-Heviside units.
We also need to consider the plasma effect and the Cotton-Mouton effect in general.
The resultant equation reads
\begin{align*}
  i\frac{d }{dz}\Psi\ =\ \bm{M} \Psi\ ,
\end{align*}
where the mixing matrix $\bm{M}$ can be defined as follows:
\begin{equation}
    \bm{M}\ \equiv\ 
     \left(
       \begin{array}{ccc}
        \Delta_a & \Delta_M & 0 \\
        \Delta_M & \Delta_{\parallel} &0\\
        0 & 0& \Delta_{\perp}
       \end{array}
     \right)
   \ \equiv 
     \left(
      \begin{array}{ccc}
        \bm{M}_{\parallel} &\bm{0} \\
        \bm{0}& \Delta_{\perp} \\
      \end{array}
     \right )\ ,\qquad
        \bm{M}_{\parallel}\ \equiv \ 
     \left(
      \begin{array}{ccc}
        \Delta_a & \Delta_M \\
        \Delta_M & \Delta_{\parallel} \\
      \end{array}
     \right)\ .
\end{equation}
Now, the photon can have an effective mass $\Delta_*\, \ (\,*\,=\,\parallel,\perp)$ due to additional effects
\begin{equation*}
  \Delta_*\ =\ \Delta_{QED} + \Delta_{CM} + \Delta_{plasma} \,.
\end{equation*}
Here, $\Delta_{QED}$ is the effect of vacuum polarization with
\begin{align}
  \begin{split}
    &\Delta^\parallel_{QED}\ \equiv\ -\frac{7}{2}\,\omega\,\rho\,B^2_T,\ \ \Delta^{\perp}_{QED}\ \equiv\ -2\,\omega\,\rho\,B^2_T\ , \\
    &\rho B^2_T\ \equiv\ \frac{4}{45m_e^4}\left(\frac{e^2}{4\pi}\right)^2B^2_T\ =\ \frac{1}{45\pi}\left(\frac{e^2}{4\pi}\right)^2\left(\frac{B_T}{B_{cr}} \right)^2\ ,\\
    &B_{cr}\ \equiv\ \frac{m_e^2}{e}\ =\ 4.42\times10^{13}\,\rm{G}\ ,
  \end{split}
\end{align}
and $\Delta_{plasma}$ denotes plasma effect
\begin{align}
  \begin{split}
    &\Delta_{plasma}\ =\ \frac{\omega^2_{p}}{2\omega}\ ,\\
    &\omega^2_{p}\ \equiv\ e^2\frac{n_e}{m_e}\ ,
  \end{split}
\end{align}
where $n_e$ is the electron density.
The Cotton-Mouton effect $\Delta_{CM}$ will be ignored in this paper.

\subsection{Single Domain Model}

In this simplest case, the axion can mix with the only parallel component $A_\parallel$ and one can reduce the mixing matrix to $2\times2$ matrix.
\begin{equation}\label{constant}
  i \ \frac{d}{dz} 
    \left(
     \begin{array}{c}
       a(z) \\
       A_\parallel(z) \\
     \end{array}
    \right)
  =
    \left(
      \begin{array}{ccc}
        \Delta_a & \Delta_M \\
        \Delta_M & \Delta_\parallel \\
      \end{array}
    \right)
    \left(
      \begin{array}{c}
        a(z) \\
        A_\parallel(z) \\
      \end{array}
    \right)\ .
\end{equation}

Here, we evaluate the photon-axion conversion probability.
To this end, we diagonalize $\bm{M}_\parallel$ in (\ref{constant}) and obtain eigenvalues $\lambda_{\pm}$
\begin{equation}
   \lambda_{\pm}\ \equiv\ \frac{(\Delta_\parallel+\Delta_a)\pm \sqrt{(\Delta_a-\Delta_\parallel)^2+(2\Delta_M)^2}}{2}\ .
\end{equation}
We introduce an orthogonal matrix $\bm{O}$, which diagonalize $\bm{M}_\parallel$
\begin{equation} \label{matrix}
 \bm{O M_{\parallel}O}^\dag\ =\ \left(
      \begin{array}{ccc}
        \lambda_+& 0\\
       0 & \lambda_- \\
      \end{array}
    \right),\ \ \ \ \ 
 \bm{O}\,\equiv\,
   \left(
      \begin{array}{ccc}
        \cos \theta& \sin \theta \\
       -\sin \theta & \cos \theta \\
      \end{array}
    \right),
\end{equation}
where $\theta$ is the mixing angle.
Defining $\tilde{\Psi}\equiv\bm{O}\,\Psi$, we can solve (\ref{constant}) as
\begin{equation}
  \tilde{\Psi}_i(z)\ =\ \tilde{\Psi}_i(z_0)\,e^{-i \lambda_i z}\ .
\end{equation}
Thus, we obtain
\begin{equation}
  \Psi_i(z)\ =\ \sum_{j=1}^2 O^{\dag}_{ij} \tilde{\Psi}_j
\ =\ \sum_{j=1}^2 O^{\dag}_{ij}[\bm{O} \Psi(0)]_j e^{-i \lambda_j z}\ ,
\end{equation}
where $\lambda_1\equiv \lambda_+$, and $\lambda_2\equiv \lambda_-$.
Thus, we obtain
\begin{align}
\label{ans_const}
\begin{split}
   a(z)\ &=\ \left( \cos^2 \theta e^{-i \lambda_+ z}+ \sin^2 \theta e^{-i \lambda_- z} \right) a(0)
   + \cos\theta\sin\theta\left(e^{-i \lambda_+ z}-e^{-i \lambda_- z}\right) A_{\parallel}(0) \ ,\\
   A_\parallel(z)\ &=\ \cos\theta\sin\theta\left(e^{-i \lambda_+ z}-e^{-i \lambda_- z}\right)  a(0)
   + \left( \sin^2 \theta e^{-i \lambda_+ z}+ \cos^2 \theta e^{-i \lambda_- z} \right) A_{\parallel}(0)
    \ .
\end{split}
\end{align}
Assuming $a(0) = 0$ and $A_{\parallel}(0) = 1$, we obtain the conversion probability of photon into axion as
\begin{align}\label{P_01}
  \begin{split}
    P_0(\gamma \to a)\ &\ =\ (\sin 2\theta)^2\sin^2 \left(\frac{\sqrt{(\Delta_a-\Delta_\parallel)^2+(2\Delta_M)^2}}{2}z\right)\ .
  \end{split}
\end{align}
Here we introduce oscillation length $\Delta_{osc}^{-1}$
\begin{equation}
  \Delta_{osc}\ \equiv\ \lambda_+-\lambda_-\ =\ \sqrt{(\Delta_a-\Delta_\parallel)^2+(2\Delta_M)^2}\ .
\end{equation}
Noticing the relation 
\begin{equation*}
  \sin 2\theta\ =\ \frac{2\Delta_M}{\Delta_{osc}}\ ,
\end{equation*}
we can rewrite conversion probability (\ref{P_01}) as
\begin{equation}\label{P_02}
  P_0(\gamma \to a)\ =\ (\Delta_M z)^2\,\frac{\sin^2 \left(\dfrac{\Delta_{osc}}{2} z\right)}{\left(\dfrac{\Delta_{osc}}{2}z\right)^2}\ .
\end{equation}
The photon-axion conversion is most effective when the strong coupling condition $\Delta_M \gg |\Delta_a-\Delta_\parallel|$ is satisfied
\begin{equation}
  \tan 2 \theta_s\ =\  \frac{2\Delta_M}{\Delta_a-\Delta_\parallel} \gg 1 
\end{equation}
This corresponds to
\begin{equation}  
  \theta_s \ \sim\ \frac{\pi}{4},\ \   \Delta_{osc,s}\ \sim\ 2\Delta_M\ . 
\end{equation}  
At this time, the conversion probability can be written in a simpler form:
\begin{equation}
  P_{0s}(\gamma \to a)\ \sim\ (\Delta_M z)^2\ .
\end{equation}

Only one of photon components can mix with the axion, so the photon-axion conversion can affect polarization of the photon.
The polarization is described by the Stokes parameters $(I,Q,U,V)$ defined as
\begin{equation}
    \begin{cases}
      I(z) &\equiv \ A_\parallel (z)A_\parallel^*(z)+A_\perp (z)A_\perp^*(z) \ ,\\
      Q(z) &\equiv \ A_\parallel (z)A_\parallel^*(z)-A_\perp (z)A_\perp^*(z) \ ,\\
      U(z) &\equiv \ A_\parallel(z)A_\perp^*(z)+A_\perp(z)A_\parallel^*(z) \ ,\\
      V(z) &\equiv i\left( \ A_\parallel(z)A_\perp^*(z) - A_\perp(z)A_\parallel^*(z) \right) \ .
   \end{cases}
\end{equation}
Note that $I(z)$, $V(z)$ are invariant under the coordinate transformation, but $Q(z)$, $U(z)$ depend on the coordinate system.
Using the Stokes parameters, we can define the degree of circular polarization
\begin{equation}
    \Pi_C\ =\ \frac{|V(z)|}{I(z)}\ ,
\end{equation}
and the degree of linear polarization
\begin{equation}
  \Pi_L\ =\ \frac{\sqrt{Q^2(z)+U^2(z)}}{I(z)}\ ,
\end{equation}
where $I(z)$ is the intensity of the photon.
They satisfy the following condition:
\[I^2(z)\ =\ Q^2(z)+U^2(z)+V^2(z)\ .\]

In the later section, we focus on polarization induced by the photon-axion conversion.
For the sake of simplicity, we assume $\Delta_{\parallel} = \Delta_{\perp} = \Delta_{plasma}$.
In this case the behavior of polarization depends only on $(\Delta_{a} - \Delta_{plasma})$.
The reason is as follows.
The Stokes parameters do not change when we redefine the field as $\Psi_{i}(z)\,e^{i\Delta_{plasma}z}$.
While the mixing matrix is transformed as
\begin{equation*}
	\bm{M} =
		\begin{pmatrix}
			\Delta_a - \Delta_{plasma} & \Delta_M & 0 \\
			\Delta_M & 0 & 0\\
			0 &0 & 0
		\end{pmatrix} \ .
\end{equation*}
Thus, we can consider only the case $\Delta_{plasma} = 0$ without loss of generality.
The result in the case $\Delta_{plasma} \not= 0$ can be obtained by replacing $\Delta_{a}$ with $(\Delta_{a} - \Delta_{plasma})$.
We can evaluate degree of circular polarization analytically,
\begin{equation}
  \Pi_C(z)\ =\ \left|\frac{U_0 (\Delta_M^2/ \Delta_{osc})}{I_0-(I_0+Q_0)(\Delta_M/\Delta_{osc})^2\left[1-\cos(\Delta_{osc}z)\right]}
\left[ \frac{\sin (\lambda_+z)}{\lambda_+}-\frac{\sin(\lambda_-z)}{\lambda_-} \right]\right|\ ,
\end{equation}
where $I_0$, $Q_0$, $U_0$ represent the initial Stokes parameters and we set $V_0=0$.
We can also evaluate other variables with the same assumption as
\begin{align}
  I (z) &= I_0 - (I_0 + Q_0)  \left(\frac{\Delta_M}{ \Delta_{osc}}\right)^2 \left[ 1- \cos \Delta_{osc} z \right] \ , \\
  Q(z) &=  Q_0 - (I_0 + Q_0)  \left(\frac{\Delta_M}{ \Delta_{osc}}\right)^2 \left[ 1- \cos \Delta_{osc} z \right] \ , \\
  U(z) &=   \frac{U_0 \Delta_M^2}{ \Delta_{osc}}  \left[ \frac{\cos (\lambda_+z)}{\lambda_+}-\frac{\cos(\lambda_-z)}{\lambda_-} \right]
  \ .
\end{align}

\subsection{Conventional Model}

It is known that the universe is magnetized.
As we have seen, the photon and the axion can mix with each other in the presence of external magnetic field.
The conventional setup in the astrophysical cosmological application is as follows.
One divide the line of sight into domains of equal length $s$.
Within each domain, the direction of magnetic field is constant, but between domains its direction randomly changes.
The photon from cosmological distance traverses these domains on its way to the earth.
The strength of magnetic field is assumed to be the same in all domains.
We refer to this scenario as the conventional model in this paper.
Here, we give the total conversion probability and the variance of polarization after passing $N$-domains in the conventional model~\cite{Ia:4}.

\begin{figure}[h]
\includegraphics[width=8cm]{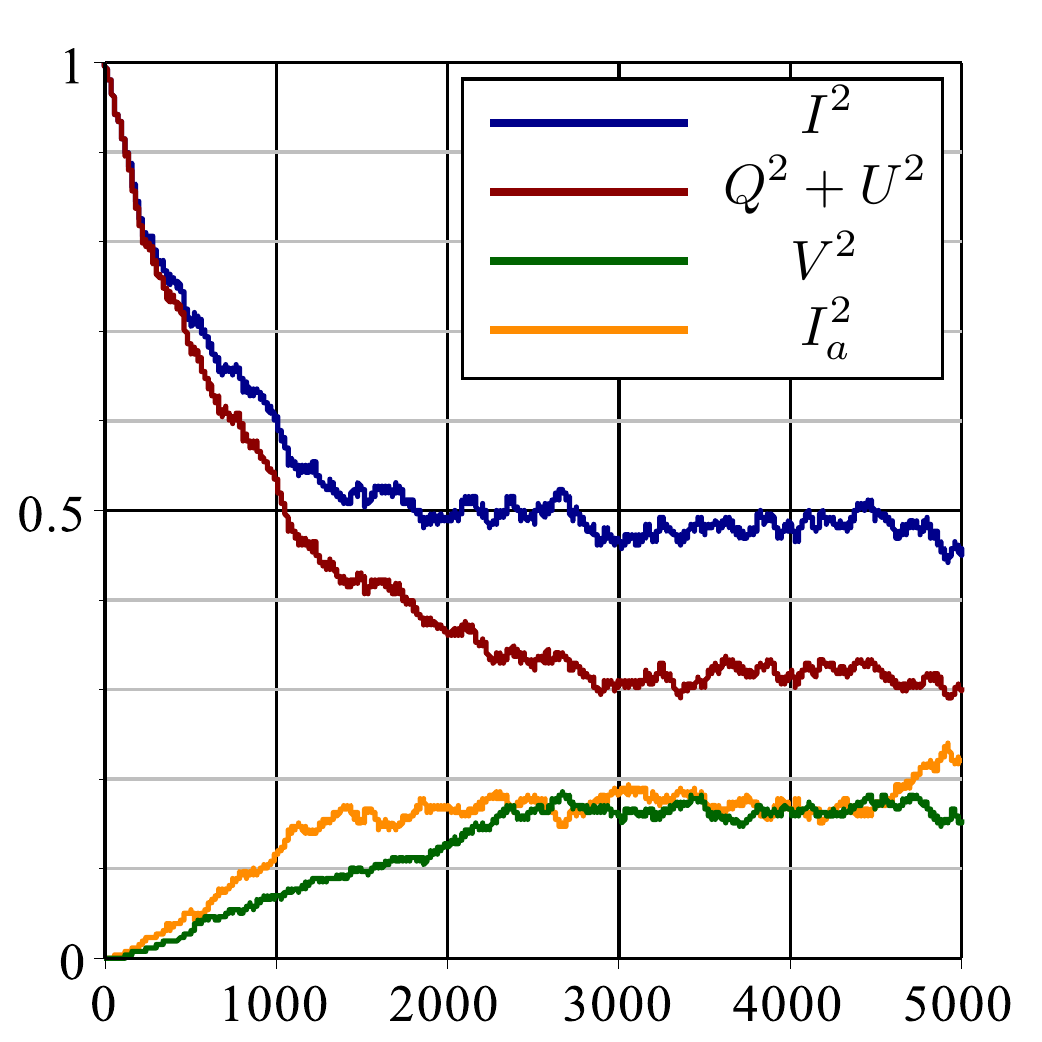}
\caption{The evolution of mean-square values of the polarization and intensity throughout 5000 domains for the conventional model.
We took the average of 100 times trial.
Since the average values of the polarizations are zero, $Q^2$, $U^2$, and $V^2$ are the variances.
We set $\Delta_a=1.5\times10^{-2}~{\rm Mpc}^{-1}$, $\Delta_M=3.0\times10^{-2}~{\rm Mpc}^{-1}$.	
}
\label{fig:discrete}
\end{figure}

We describe the length of one domain as $s$, so the conversion probability within a single domain is
\begin{equation}
  P_0(\gamma \to a)\ =\ (\Delta_M s)^2\,\frac{\sin^2 \left(\dfrac{\Delta_{osc}}{2} s\right)}{\left(\dfrac{\Delta_{osc}}{2}s\right)^2}\ .
\end{equation}
One can use any coordinate system whichever you want as long as one pays attention to the invariant quantity, such as $I$, $\Pi_C$, $\Pi_L$.
Within a single domain, we have already known the solutions (\ref{ans_const}).
We define the $n$-th domain as $z_{n-1}<z\leqq z_n$.
The angle between $\bm{B}_T$ in the $n$-th domain and the $x$-axis in the fixed $x$-$y$ coordinate is denoted as $0\leqq\varphi_n\leqq2\pi$, and we set $\varphi_0 = 0$.
Now, we can evaluate the total conversion probability after passing $N$-domains.
The relationship between the ($n-1$)-th domain and $n$-th domain is given by
\begin{align}
\label{recursion}
   a^n (z_n) \ &=\ \left( \cos^2 \theta e^{-i \lambda_+ s}+ \sin^2 \theta e^{-i \lambda_- s} \right) a(z_{n-1})\nonumber \\
  &   + \cos\theta\sin\theta\left(e^{-i \lambda_+ s}-e^{-i \lambda_- s}\right) 
      \left( \cos \gamma_n A_{\parallel}^{n-1} (z_{n-1}) + \sin \gamma_n A_\perp^{n-1} (z_{n-1}) \right) \ ,\\
   A_\parallel^n (z_n)\ &=\ + \cos\theta\sin\theta\left(e^{-i \lambda_+ s}-e^{-i \lambda_- s}\right)  a(z_{n-1}) \nonumber\\
  & + \left( \sin^2 \theta e^{-i \lambda_+ s}+ \cos^2 \theta e^{-i \lambda_- s} \right) 
  \left( \cos \gamma_n A_{\parallel}^{n-1} (z_{n-1}) + \sin \gamma_n A_\perp^{n-1} (z_{n-1}) \right)  \ ,\\
   A_\perp^n (z_n) \ &=  - \sin \gamma_n A_{\parallel}^{n-1} (z_{n-1}) + \cos \gamma_n A_\perp^{n-1} (z_{n-1}) 
    \ .
\end{align}
where $\gamma_n\equiv \varphi_n-\varphi_{n-1}$.
The intensity of the photon and the axion at the end of the $n$-th domain is represented by quantities of the ($n-1$)-th domain
\begin{align}
  \begin{split}
    I(z_n)\ &=\ P_0|a(z_{n-1})|^2+(1-P_0\cos^2\gamma_n )|A_\parallel^{n-1}(z_{n-1})|^2+(1-P_0\sin^2\gamma_n )|A_\perp^{n-1}(z_{n-1})|^2+\cdots\ ,\\
    I_a(z_n)\ &=\ (1-P_0)|a(z_{n-1})|^2+\cos^2\gamma_n P_0|A_\parallel^{n-1}(z_{n-1})|^2+\sin^2\gamma_n P_0|A_\perp^{n-1}(z_{n-1})|^2+\cdots \ ,
  \end{split}
\end{align}
where the dots represent terms which are proportional to $\cos \gamma_n$, $\sin \gamma_n$, or $\cos \gamma_n \sin \gamma_n$.
We assume that the angles $\gamma_n$ are randomly chosen.
Because of the randomness, $\cos^2 \gamma_n$ and $\sin^2 \gamma_n$ can be replaced by their average value $1/2$, and the other terms are averaged to zero.
Then we obtain the recursion relation
\begin{align}\label{w}
  \begin{split}
    \left(
     \begin{array}{c}
       I(z_n) \\
      I_a(z_n) \\
     \end{array}
    \right)
   \ &\equiv\ \bm{W}\left(
      \begin{array}{c}
       I(z_{n-1})\\
      I_a(z_{n-1}) \\
      \end{array}
    \right)\ ,
 \end{split}
\ \ \  \bm{W}\ \equiv\ \left(
      \begin{array}{ccc}
       1-\frac{1}{2}P_0&P_0 \\
       \frac{1}{2}P_0 & 1-P_0 \\
      \end{array}
    \right).
\end{align}
Now we evaluate the eigenvalues $w_\pm$ of the matrix $\bm{W}$.
\begin{equation}
  w{_\pm}\ \equiv\ \frac{1}{2}\left[\left(2-\frac{3}{2}P_0\right)\pm\frac{3}{2}P_0\right] \ .
\end{equation}
Then, from Eq.~(\ref{w}), we can deduce the following
\begin{equation}
    \left(
     \begin{array}{c}
        I(z_n) \\
       I_a(z_n) \\
     \end{array}
    \right)
    =\ \frac{1}{3}\left(
      \begin{array}{ccc}
       2+\left(1-\frac{3}{2}P_0\right)^{n}&2-2\left(1-\frac{3}{2}P_0\right)^{n} \\
       1-\left(1-\frac{3}{2}P_0\right)^{n}&1+2\left(1-\frac{3}{2}P_0\right)^{n} \\
      \end{array}
    \right)
    \left(
      \begin{array}{c}
       I(z_0)\\
       I_a(z_0) \\
      \end{array}
    \right)  
    \ .
\end{equation}
We describe the length of one domain as {\it s}, so $n=z/{\it s}$ gives a number of domains.
Using the relation
$
  e^x\ =\ \lim_{n \to \infty} \left(1+x/n\right)^n\ ,
$
we arrive at
\begin{align*}
  \left(1-\frac{3}{2}P_0 \right)^{n}\ =\ \left(1+\frac{-\frac{3}{2}P_0\frac{z}{s}}{\frac{z}{s}}\right)^{n}\ =\ \left(1+\frac{-\frac{3P_0}{2s}z}{n}\right)^{n}\ \xrightarrow[n \to \infty]{}\ e^{-\frac{3P_0}{2s}z}\ .
\end{align*}
Eventually, we obtain the following expressions
\begin{align}
  \begin{split}
    &I(z)\ =\ I(z_0)-P_{\gamma\to a}[I(z_0)-2I_a(z_0)]\ ,\\
    &I_a(z)\ =\ I_a(z_0)+P_{\gamma\to a}[I(z_0)-2I_a(z_0)]\ ,
  \end{split}
\end{align}
with
\begin{equation}
  P_{\gamma \to a}\ =\ \frac{1}{3}\left[1-e^{-\frac{3P_0}{2s}z}\right]\ .
\end{equation}
In the limit $n=z/s \to \infty$, the conversion probability saturates so that on average one third of all photons converts to axions.
To find the same limit for other quantities, we introduce density matrix and Stokes parameters $(I,Q,U,V)$
\begin{align}
  \bm{\rho}(z)\ &\equiv\ 
    \left(\begin{array}{c}
       a(z)\\
       A_\parallel(z)\\
       A_\perp(z)
    \end{array}\right)
  \otimes \left(a^*(z)\ A^*_\parallel(z)\ A^*_\perp(z)\right) \nonumber\\
  \ &=\  \left(
    \begin{array}{ccc}
      I_a (z)  &  \dfrac{K(z)-iL(z)}{2}  &  \dfrac{M(z)-iN(z)}{2} \\
       \dfrac{K(z)+iL(z)}{2} & \dfrac{I(z)+Q(z)}{2} &\dfrac{U(z)-iV(z)}{2} \\
       \dfrac{M(z)+iN(z)}{2} &\dfrac{U(z)+iV(z)}{2} & \dfrac{I(z)-Q(z)}{2}
    \end{array}
    \right)\ ,
\end{align}
where we defined $2 a\,A_\parallel^* =K-iL$ and $2 a\,A_\perp^* = M-iN$.
We can repeat the same analysis by including polarization degrees of freedom.
Then, we obtained the asymptotic values
\begin{equation}
I_a = \frac{1}{3} \ , I =\frac{2}{3}  \ , Q = U = V = K = L= M = N = 0 \label{limit}\ .
\end{equation}
Thus, the polarization vanishes on average.
However, what we need to evaluate is the variance of polarization.
To this aim, it is useful to notice that the density matrix obeys
\[i\frac{d \bm{\rho}(z)}{dz}\ =\ [\bm{M},\bm{\rho}(z)]\ .\]
If we define the transfer function $\bm{T}(z,z_0)$ with initial condition $\bm{T}(z_0,z_0)\,=\,1$, then we can formally solve the dynamics as
\begin{equation}
  \bm{\rho}(z)\ =\ \bm{T}(z,z_0)\,\bm{\rho}(z_0)\,\bm{T}^\dag(z,z_0)\ .
\end{equation}
Remarkably, $\bm{\rho}^2(z)$ also obeys the same equation
\begin{equation}
  \bm{\rho}^2 (z)\ =\ \bm{T}(z,z_0)\,\bm{\rho}^2 (z_0)\,\bm{T}^\dag(z,z_0)\ .
\end{equation}
This implies the corresponding components of $\bm{\rho}^2(z)$ have the same limit values as those in (\ref{limit}).
From the explicit calculation
\begin{align}
  \bm{\rho}^2 (z)\ 
  = \  \left(
    \begin{array}{ccc}
      I^2_a +\frac{K^2 +L^2 + M^2 + N^2}{4}  &  {\rm cross \ terms} &{\rm cross \ terms} \\
      {\rm cross \ terms}& \frac{K^2 +L^2 +(I+Q)^2 +U^2 + V^2}{4} &{\rm cross \ terms} \\
      {\rm cross \ terms}&{\rm cross \ terms} & \frac{M^2 +N^2 +(I-Q)^2 +U^2 + V^2}{4} 
    \end{array}
    \right)\ ,
\end{align}
we obtain the asymptotic values
\begin{align}
& I^2_a +\dfrac{K^2 +L^2 + M^2 + N^2}{4} = \frac{1}{3} \ , \\
& \frac{I^2+Q^2 +U^2 + V^2}{2}+\frac{K^2 +L^2 + M^2 + N^2}{4} =\frac{2}{3}  \ , \\
& K^2 +L^2 - M^2 -  N^2   =0 \ , \\
&  {\rm cross \ terms} =0 \ ,
\end{align}
where we ignored average of the cross terms such as $IQ$ because of the statistical independence of these variables.
Assuming equi-partition for $I^2_a,Q^2,U^2,V^2,K^{2},L^2,M^2,N^2$ and taking into account the relation $I^2 =Q^2 +U^2 +V^2$, we can conclude
\begin{equation}
I^2_a = \frac{1}{6} \ , I^2 =\frac{1}{2}  \ , Q^2 =\frac{1}{6} \ , U^2 =\frac{1}{6} \ , V^2 =\frac{1}{6} \ .
\end{equation}
We confirmed theses results numerically in Fig.~\ref{fig:discrete}.

Now, we can analytically calculate the degree of circular polarization
\begin{equation}
    \Pi_C\ =\ \sqrt{\frac{V^2}{I^2} }= 0.577\ ,
\end{equation}
and the degree of linear polarization
\begin{equation}
  \Pi_L\ =\ \sqrt{\frac{Q^2+U^2}{I^{2}}}= 0.816\ .
\end{equation}
The latter shows a good agreement with numerical result in \cite{polarization:4}.

\section{Photon-Axion Conversion in Helical Domain Model}\label{3}

The actual configuration of cosmological magnetic fields is not well known.
If we assume that magnetic fields have primordial origin, then the magnetic fields can have helicity.
Indeed, the equations of motion of photon-axion conversion in helical single domain model have been already analyzed~\cite{helical:3}.
However, there are several possibilities how to connect its domain.
First, we review equations of motion within a helical domain.
Next, we connect domain in a different way from Wang and Lai model~\cite{helical:3} and
we performe detailed analysis of the photon-axion conversion process in our model and conventional model.
Our motivation for this study is to understand the magnetic field configuration dependence on the dynamics of polarization in the photon-axion conversion process and obtain new constraints on the model parameters by utilizing the results.

\subsection{Helical Single Domain Model}

In the  conventional model, the background magnetic field is constant in each domain, namely,
\[ \bm{B}_T \ =\ B_T\,\hat{e}_\parallel\ . \]
In the helical model, even within a single domain, the direction of magnetic field changes in the fixed $x$-$y$ coordinate system
\[\overline{\bm{B}}_T \ =\ B_T\cos(\alpha z)\,\hat{e}_x + B_T\sin (\alpha z)\,\hat{e}_y\ .\]

The linearized equations of motion are given by
\begin{equation}
  \begin{cases}
   \left[(\omega+i\partial_z)-\dfrac{m_a^2}{2\omega}\right]a\ =\ -\dfrac{1}{2}g_{a\gamma\gamma}\,\omega B_T \left[A_x\cos(\alpha z)+A_y\sin(\alpha z)\right]\ , \\
  (\omega + i \partial_z) A_x\ =\ -\dfrac{1}{2}g_{a\gamma\gamma}\,B_T\,\cos(\alpha z)\,a \ ,\\
  (\omega + i \partial_z) A_y\ =\ -\frac{1}{2}g_{a\gamma\gamma}\,B_T\sin(\alpha z)\,a\ ,
  \end{cases}
\end{equation}
where we used $\Box\,\simeq\,2\omega(\omega+i\partial_z)$.
Shifting the phase in the same way as before, we arrive at the following equation:
\begin{equation}\label{xy}
  i\frac{d}{d z}
    \left(
     \begin{array}{c}
       a(z) \\
       A_x(z) \\
       A_y(z)
     \end{array}
    \right)
  = 
    \left(
     \begin{array}{ccc}
        \Delta_a& \Delta_M \cos (\alpha z) & \Delta_M \sin (\alpha z) \\
        \Delta_M \cos (\alpha z) & \Delta_{xx} & \Delta_{xy}\\\
        \Delta_M \sin (\alpha z) & \Delta_{yx} &\Delta_{yy}
     \end{array}
    \right)
    \left(
     \begin{array}{c}
       a (z)\\
       A_x(z)\\
       A_y(z)
     \end{array}
    \right)\ ,
\end{equation}
with
\begin{align*}
  \begin{split}
    \Delta_{xx}\ &=\ \Delta_\parallel \cos^2(\alpha z)+\Delta_\perp \sin^2 (\alpha z)\ ,\\
    \Delta_{xy}\ &=\ \Delta_{yx}\ =\ (\Delta_{\parallel}-\Delta_{\perp}) \cos (\alpha z)\sin (\alpha z)\ ,\\
    \Delta_{yy}\ &=\ \Delta_{\parallel}\sin^2(\alpha z)+\Delta_{\perp}\cos^2(\alpha z)\ .
  \end{split}
\end{align*}
Here, $\Delta_* (\,*\,=\,\parallel,\perp)$ represent additional effects as we have already seen
\[\Delta_*\ =\ \Delta_{QED} +  \Delta_{plasma} \,,\]
and $\Delta_a$, $\Delta_M$ are defined similarly
\[\Delta_a\ =\ \frac{m_a^2}{2\omega}\,,\ \ \ \Delta_M\ \equiv\ -\frac{1}{2}g_{a\gamma\gamma}B_T.\]

The QED effect should be modified in helical magnetic fields.
However, since the modifications, which come from derivatives of $\sin(\alpha z)$ and $\cos(\alpha z)$, are proportional to $\alpha / k \sim \alpha / \omega \ll 1$, we can neglect them at the leading order.
Hence, we can use the same form of $\Delta_{QED}$ as that in the constant magnetic field model.

We perform the unitary transformation $\bm U$ and change coordinate basis
\begin{equation*}
  \left(
   \begin{array}{c}
       a (z)\\
       A_\parallel(z)\\
       A_\perp(z)
   \end{array}
        \right)
   \ =\  \left(
   \begin{array}{ccc}
        1 & 0& 0  \\
        0  & \cos (\alpha z) & \sin (\alpha z)\\
        0  & - \sin (\alpha z) & \cos (\alpha z)
     \end{array}
    \right)
    \left(
   \begin{array}{c}
       a (z)\\
       A_x(z)\\
       A_y(z)
     \end{array}
    \right)
   \ \equiv\ \bm{U}
   \left(
     \begin{array}{c}
       a (z)\\
       A_x(z)\\
       A_y(z)
     \end{array}
    \right)\ ,
\end{equation*}
from $x$-$y$ coordinate to $\parallel$-$\perp$ coordinate.
We can rewrite Eq.~(\ref{xy}) as
\begin{equation}\label{continuous}
  i\frac{d}{d z}
    \left(
     \begin{array}{c}
       a(z) \\
       A_\parallel(z) \\
       A_\perp(z)
     \end{array}
    \right)
  = 
    \left(
     \begin{array}{ccc}
        \Delta_a & \Delta_M & 0 \\
        \Delta_M & \Delta_{\parallel} & i \alpha\\
        0 & -i\alpha & \Delta_{\perp}
     \end{array}
    \right)
    \left(
     \begin{array}{c}
       a (z)\\
       A_\parallel(z)\\
       A_\perp(z)
     \end{array}
    \right)\ .
\end{equation}
Thus, the mixing matrix in the continuous helical magnetic field $\overline{\bm{M}}$ is deduced as follows:
\begin{align}
  \overline{\bm{M}}\ &\equiv\ 
  \left(
     \begin{array}{ccc}
        \Delta_a & \Delta_M & 0 \\
        \Delta_M & \Delta_{\parallel} & i \alpha\\
        0 & -i\alpha & \Delta_{\perp}
     \end{array}
    \right)\ .
\end{align}

\subsection{Configuration Dependence of Polarization}\label{sec3:configuration}

We investigate the effect of the photon-axion conversion on the polarization in the presence of helical magnetic fields.
In particular, we examine to what extent the result depends on the configuration of magnetic fields.
We suppose that linearly polarized photons emitted by a source at cosmological distance converts to the axion, and {\it vise versa}, on their way to the earth.
We again divide the line of sight into domains of equal length $s=1\,{\rm Mpc}$.
We assume that $\alpha$ changes randomly from domain to domain, but the magnitude of magnetic fields is fixed for simplicity.
Formally, we can obtain the transfer function in the same way as in the conventional model.
However, in the case of the helical magnetic fields, the matrix $\overline{\bm{O}}$ which diagonalizes the mixing matrix $\overline{\bm M}$ has a complicated dependence on the parameters, so we have to resort to numerical calculations.

We take $\alpha$ as a random variable uniformly distributed within a range.
On the other hand, Wang ad Lai take $\varphi$ as a random variable, and connect them between domains continuously~\cite{helical:3}.
Note that our model is completely different from Wang and Lai model, though the dynamics within a domain is the same.
The point is that we select $\alpha$ uniformly, but they do nonuniformly.
Their choice of $\alpha$ is specific.
For example, if the direction of magnetic field of the $n$-th domain was $\pi$, then the ($n$+1)-th direction is determined in such a way as to satisfy $\alpha<0$.
This is why we try choosing $\alpha$ randomly.
\begin{figure}[h!]
\begin{center}
\includegraphics[width=10cm]{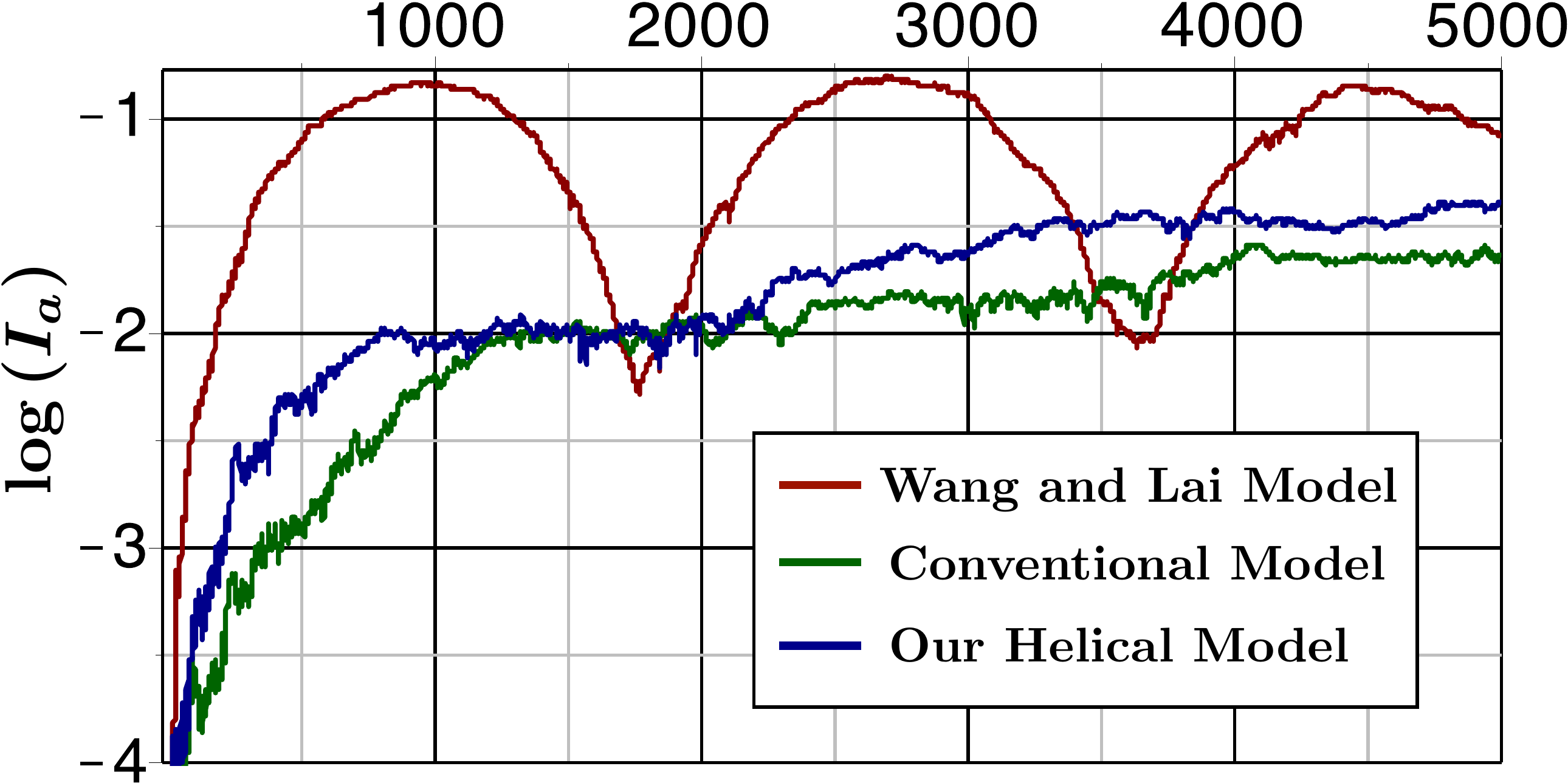}
\caption{We plotted the evolution of mean values of the axion intensity for Wang and Lai model, conventional model, and our model.  We used the same parameters as Wang and Lai paper, their Figure  2 in JCAP06(2016)006; that is, we set $\Delta_a = -7.83 \times 10^{-2}/30$, $\Delta_M = 4.63 \times 10^{-3}$ and $|\alpha| \leq \pi\,{\rm Mpc}^{-1}$ and took the average of 30 times trial.  As you can see, our model has the same statistical behavior as the conventional discreet model. }
\label{fig:w-l}
\end{center}
\end{figure}
Our model changes the coherent length, depending the range of $\alpha$.  Conventional discrete model and Wang and Lai model have coherent length in the order of Mpc.  So we should choose $\alpha$ in the range of $-\pi \sim \pi\,{\rm Mpc}^{-1}$, when we compare our model with theirs.  As you can see in Fig.~\ref{fig:w-l}, the oscillations of Wang and Lai disappear and the predictions are much closer to the conventional discontinuous magnetic field configuration model.  
The curious behavior of Wang and Lai model may be caused by their specific configuration, but to clarify this point
 is beyond the scope of this paper.

\begin{figure}[h!]
\begin{center}
\includegraphics[width=8cm]{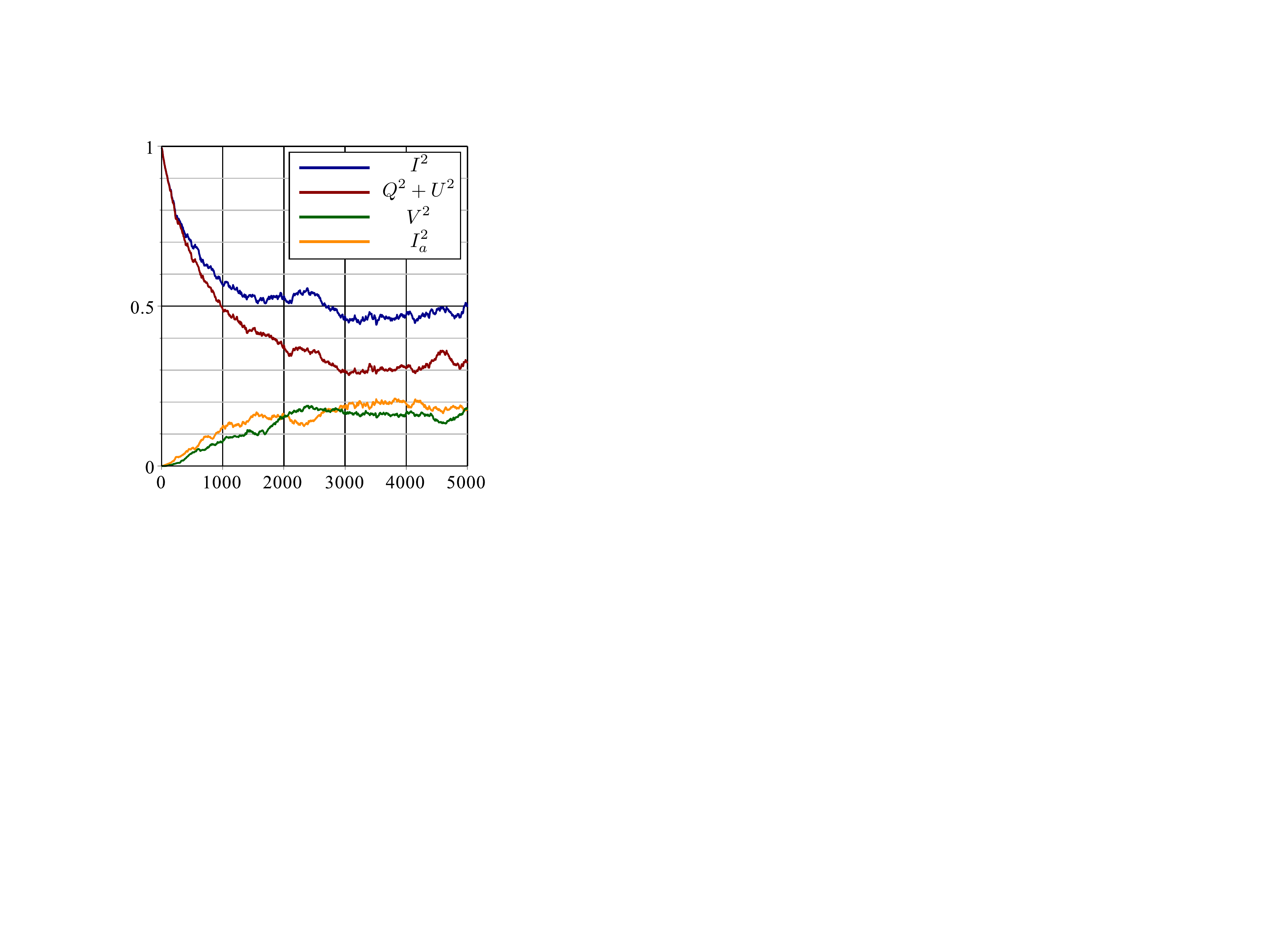}
\caption{The evolution of mean-square values of the polarization and intensity throughout 5000 domains for the helical domain model with
$|\alpha| \leq \pi \, \text{Mpc}^{-1}$.  We took the average of 100 times trial.
Since the average values of the polarizations are zero, $Q^2$, $U^2$, and $V^2$ are the variances.  We set $\Delta_a=1.5\times10^{-2}~{\rm Mpc}^{-1}$, $\Delta_M=3.0\times10^{-2}~{\rm Mpc}^{-1}$.
}
\label{fig:180}
\end{center}
\end{figure}
\begin{figure}[h]
\begin{center}
\includegraphics[width=8cm]{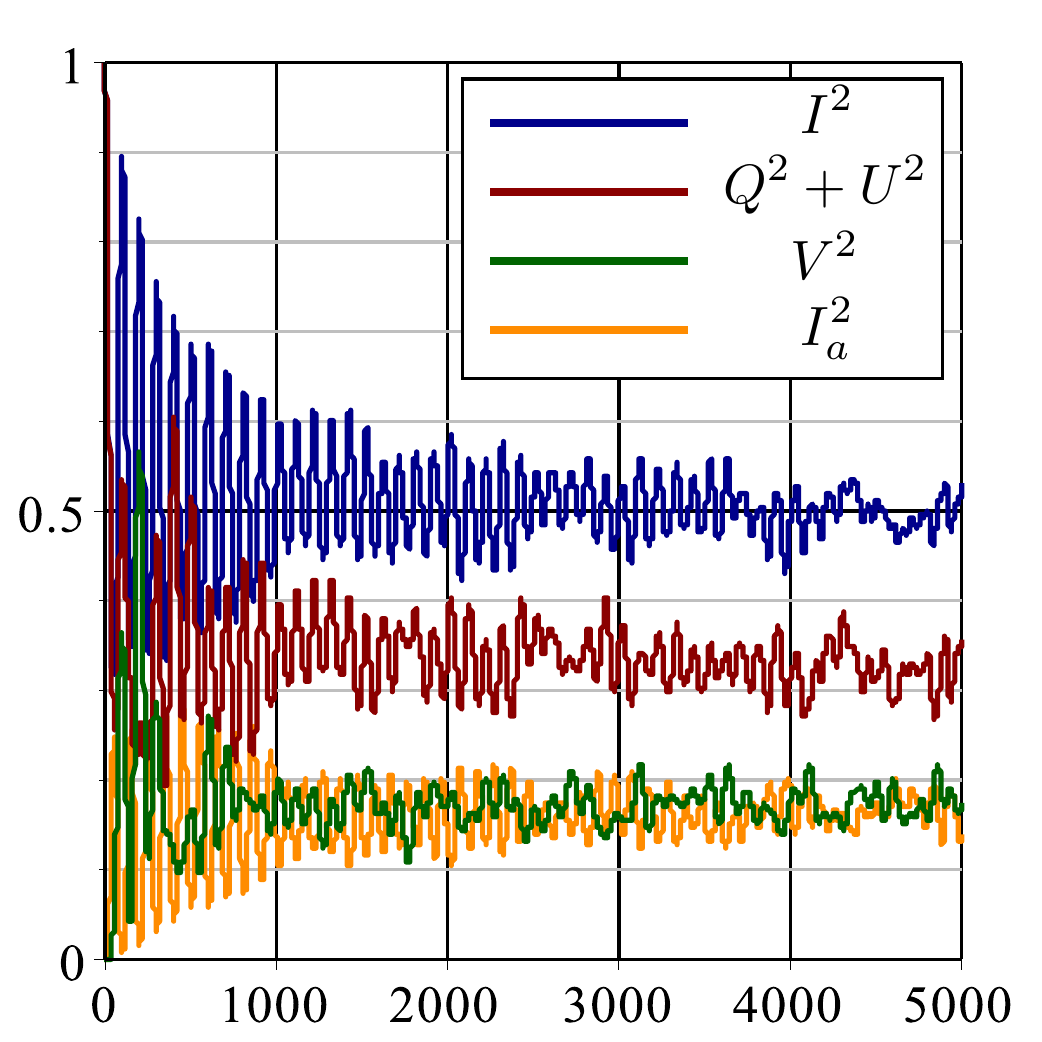}
\caption{The evolution of mean-square values of the polarization and intensity throughout 5000 domains for the helical domain model with
$|\alpha| \leq \pi / 36 \, \text{Mpc}^{-1}$.  We took the average of 100 times trial.
Since the average values of the polarizations are zero, $Q^2$, $U^2$, and $V^2$ are the variances.
In this plot, we used the parameters  $\Delta_a=1.5\times10^{-2}~{\rm Mpc}^{-1}$, $\Delta_M=3.0\times10^{-2}~{\rm Mpc}^{-1}$.
}
\label{fig:10}
\end{center}
\end{figure}
We solve Eq.~(\ref{continuous}) numerically, and examined the evolution of various variables.
In Fig.~\ref{fig:180}, we plotted the evolution of the mean-square values for relevant quantities for the range $|\alpha |\leq \pi \, \text{Mpc}^{-1}$.
The behavior of mean-square values are quite similar to those in the conventional model, as already seen in Fig.~\ref{fig:w-l}.
Due to the large change of helical configuration, the cancellation occurs 
and the mean-square values converge to the asymptotic values 
 in the same way as conventional model where the direction of magnetic field discretely changes.
However, when we makes the range for $\alpha$ small, the difference appears.
In Fig.~\ref{fig:10}, we plotted the evolution of the mean-square values for relevant quantities for the range $|\alpha |\leq \pi/36 \, \text{Mpc}^{-1}$.
As you can see, the convergence to the asymptotic values becomes slow and there appears the damped oscillating behavior in the early stage of evolution.
Due to the small change of helical configuration, the cancellation rarely occurs and the mean-square values oscillate in the same way as the case
$\alpha=0$.  However,  after traversing many domains, the each mean-square values end up with the asymptotic value.  
Thus, the converging behavior itself is universal.
If we further decrease the range of $\alpha$, we can see the oscillation lasts for a longer period.
In short, this situation increases the effective coherent length of the domains, so we see the same oscillation as fixed magnetic field.  However, its oscillation accompanies a damping, since magnetic field is not completely homogeneous.

From the theoretical point of view, several mechanism for producing cosmological magnetic fields have been suggested.
For the scenario that the magnetic fields are ejected from the galaxy through some process, the distribution of magnetic fields might be random.
On the other hand, in the case of inflationary generation of primordial magnetic fields, there should be coherence over the cosmological distances and the helicity of magnetic fields might be non-zero.
Furthermore, observations have not told us the actual magnetic configuration.
Therefore, we should keep the configuration dependence in our mind when we apply the photon-axion conversion to the astrophysics or cosmology.


\subsection{Astrophysical Application}\label{astro_helical}
\begin{figure}[h!]
\begin{center}
\includegraphics[width=10cm]{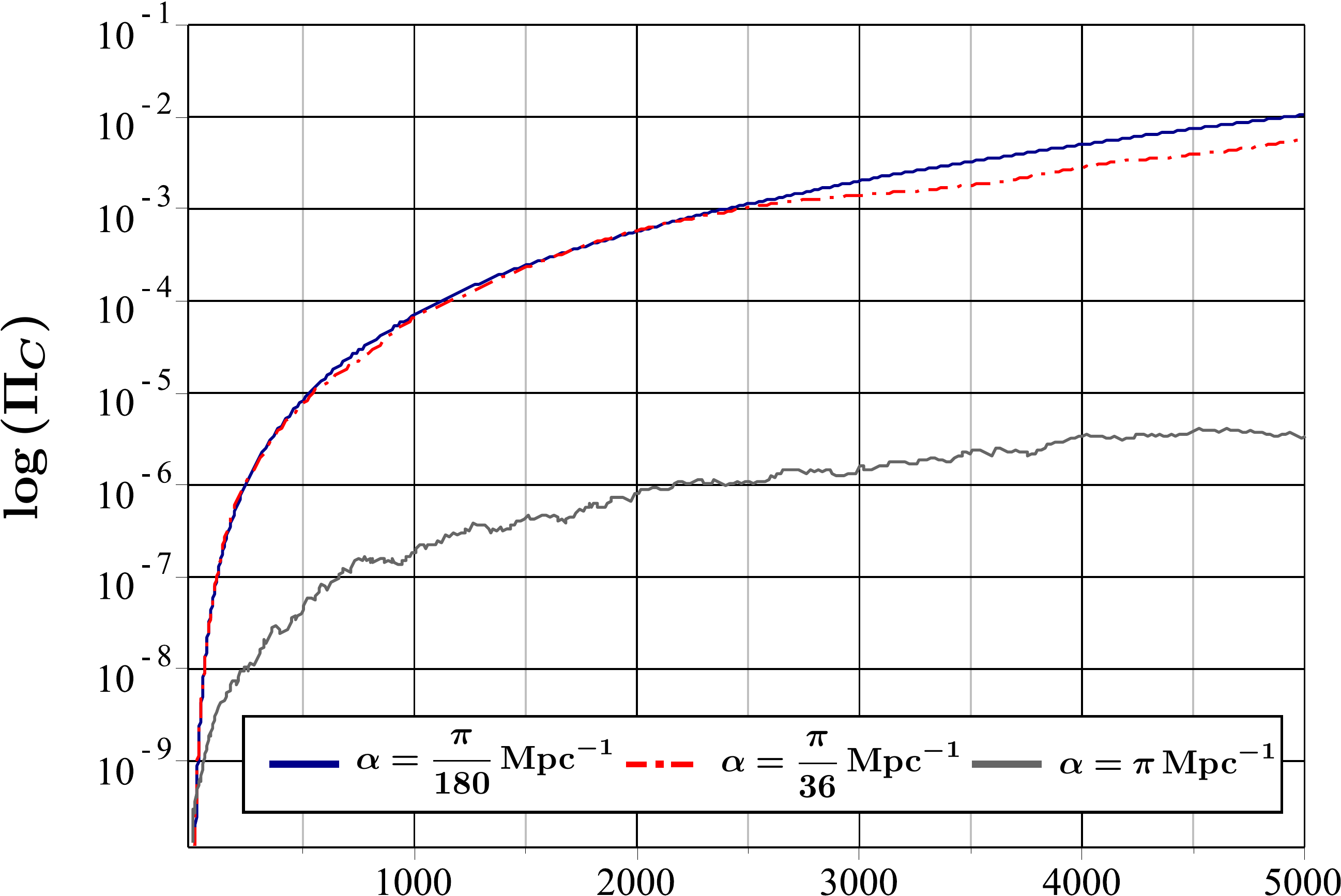}
\end{center}
\caption{The evolution of circular polarization for various ranges of $\alpha$ is depicted.
In these plots, we set $\Delta_a=7.8\times10^{-5}\,{\rm Mpc}^{-1}$ , $\Delta_M=-1.5\times10^{-4}\,{\rm Mpc}^{-1}$, and $\Delta_\parallel=\Delta_\perp=0$ and took the average of 30 times trial.
As we make the range of $\alpha$ broad, the circular polarization becomes small.}
\label{fig:comparison}
\end{figure}
Now, we are in a position to apply our results to the astrophysical situation.
For the parameters corresponding to the boundary between allowed and excluded regions, we found the mean-square values do not reach the asymptotic values.
Rather, the transient regime is relevant.
To see this, we plotted the evolution in Fig.~\ref{fig:comparison}.
As we make the range of $\alpha$ broad, the degree of circular polarization becomes small.
This is because the polarizations are almost cancelled out for a broad range of $\alpha$, as in the case of the conventional model.
Therefore, for a broad range of $\alpha$, the constraints on the parameters tend to be weak.

The effects of polarization induced by the photon-axion conversion has been widely investigated~\cite{polarization:1,polarization:2,polarization:3,polarization:4,polarization:5,polarization:6,
polarization:7,polarization:8,polarization:9,polarization:10,polarization:11,polarization:12}.
The large scale alignment of polarizations from distant quasars may be explained by conversion~\cite{polarization:2,polarization:5,polarization:6,polarization:7,polarization:8}.  
Using current polarization data, one can derive a new constraint for photon-axion coupling $g_{a\gamma\gamma}$~\cite{polarization:3,polarization:9,polarization:10,polarization:11}.  Photon-axion conversion also have another observable changes in polarization photons~\cite{polarization:1,polarization:4,polarization:12}.

In particular, a new constraint for axion coupling is reported~\cite{polarization:9,polarization:11}.
They obtained the constraint on the parameter set $(\Delta_a,\Delta_M)$ by taking into account the absence of circular polarization in the observed data.
This constraint can be translated into the constraint on $g_{a\gamma\gamma}$ or $|\bm{B}_T|$ when we fix one of them.
We postpone comparing our model with \cite{polarization:9,polarization:11} and obtaining constraints for $g_{a\gamma\gamma}$ or $|\bm{B}_T|$.
Here, we give a constraint on the parameter set $(\Delta_a,\Delta_M)$ with the helical domain model, provided the initial conditions $I_0=1$, $U_0=1$, $Q_0=0$.

Let us recall the following relevant parameters
\begin{align}
  \begin{split}
    &\Delta_a\ \equiv\ \frac{m_a^2}{2\omega}\ =\ 7.8\times10^{23}\left(\frac{m_a}{1\rm{eV}}\right)^2
     \left( \frac{100\,\rm{keV}}{\omega}\right)\,\rm{Mpc}^{-1}\ ,\\
    &\Delta_M\ \equiv\ -\frac{1}{2}g_{a\gamma\gamma}B_T\ =\ -1.5\times10^{-2} 
      \left(\frac{g_{a\gamma\gamma}}{10^{-11}\rm{GeV}^{-1}}\right)\left(\frac{B_T}{10^{-9}\rm{G}} \right)\rm{Mpc}^{-1}\ ,\\
    &\Delta_{plasma}\ \equiv\ \frac{\omega^2_{p}}{2\omega}\ =\ 1.1\times10^{-4}\left(\frac{100\,\rm{keV}}{\omega}\right)\left(\frac{n_e}{10^{-7}\,\rm{cm}^{-3}}\right)\rm{Mpc^{-1}}\ ,\\
    &\Delta_{QED}\ \sim \ \frac{\omega}{45\pi}\left(\frac{e^2}{4\pi}\right)^2\left(\frac{B_T}{B_{cr}} \right)^2\ =\ 4.1\times10^{-16}\left(\frac{\omega}{100\,\rm{keV}}\right)\left(\frac{B_T}{10^{-9}\rm{G}}\right)^2\rm{Mpc}^{-1}\ ,\\
 \end{split}
\end{align}
where we used the fact that 1\,eV $=1.57\times 10^{29}\,\rm{Mpc}^{-1}$, 1\,G $=1.95\times10^{-2}\,\rm{eV}^2$.
For the free electron number density $n_e$ at present time, we use the value $n_{e} = 10^{-7}\rm{cm}^{-3}$ throughout the paper.
This number can be obtained by assuming that all of the baryons in the universe are ionized.
We assume the mass of an axion has no relation with the coupling constant.
Absence of $\gamma$-rays from SN 1987A gives the strongest limit for the coupling constant~\cite{experiment1,experiment2,experiment5}.
We must take the plasma effect into account depending on the value of $\Delta_a$.
We consider only parameter region where we can ignore the QED effect.
The direction of magnetic field changes as $z$ varies, so the initial condition of the direction  might not have much importance for constraints on physical parameters, as long as we consider photons from cosmological distances and $\alpha$ is not too small.
For concreteness, we set $A_\parallel(z_0)=1/\sqrt{2}$, $A_\perp(z_0)=1/\sqrt{2}$.

Once we choose the range of $\alpha$, the axion mass $m_a$, and the photon energy $\omega$, then we can calculate how a linearly polarized photon is affected by the photon-axion conversion.
If the conversion produces the sizable circular polarization, it conflicts with observational results.
So we can determine the allowed region in $\Delta_a - \Delta_M$ plane by the condition that $\Pi_C$ of photon after going through 5000 domains of $s=1\,{\rm Mpc}$ does not exceed $1\%$.

\begin{figure}
\begin{center}
    \includegraphics[width=10cm]{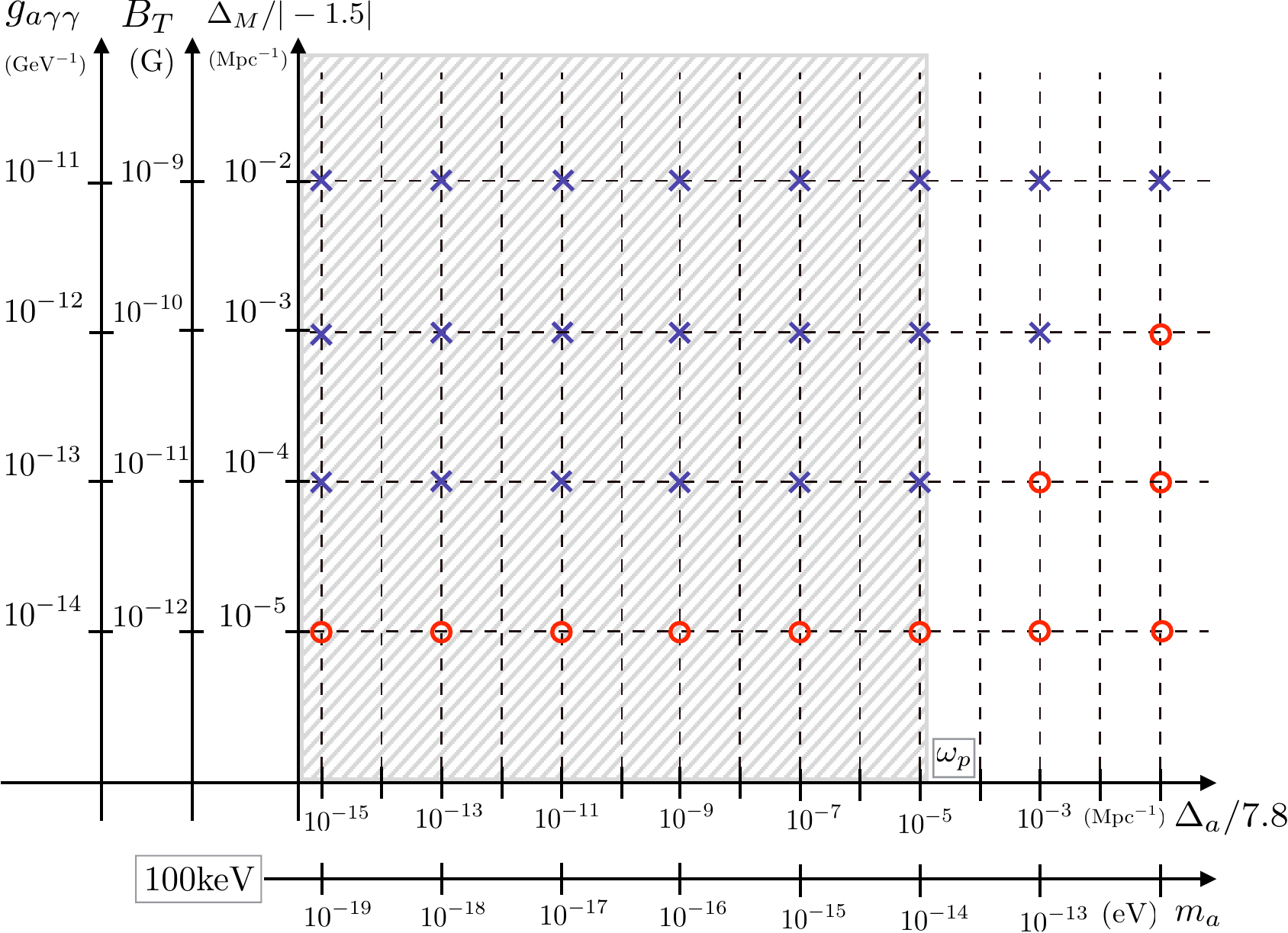}
    \caption{Constraint on IGMF using 100\,keV photons and $|\alpha |\leq \pi/180 \, \text{Mpc}^{-1}$. 
     The vertical line shows the strength of magnetic field and the horizontal one the mass of axion.
    The hatching area shows a region with the plasma effect. 
   We took the average of 30 times trial in order to constrain the parameter set.
    A cross mark represents excluded parameter set, and a red circle represents allowed parameter set.}
    \label{max}
\end{center}
\end{figure}

\begin{figure}
\begin{center}
    \includegraphics[width=10cm]{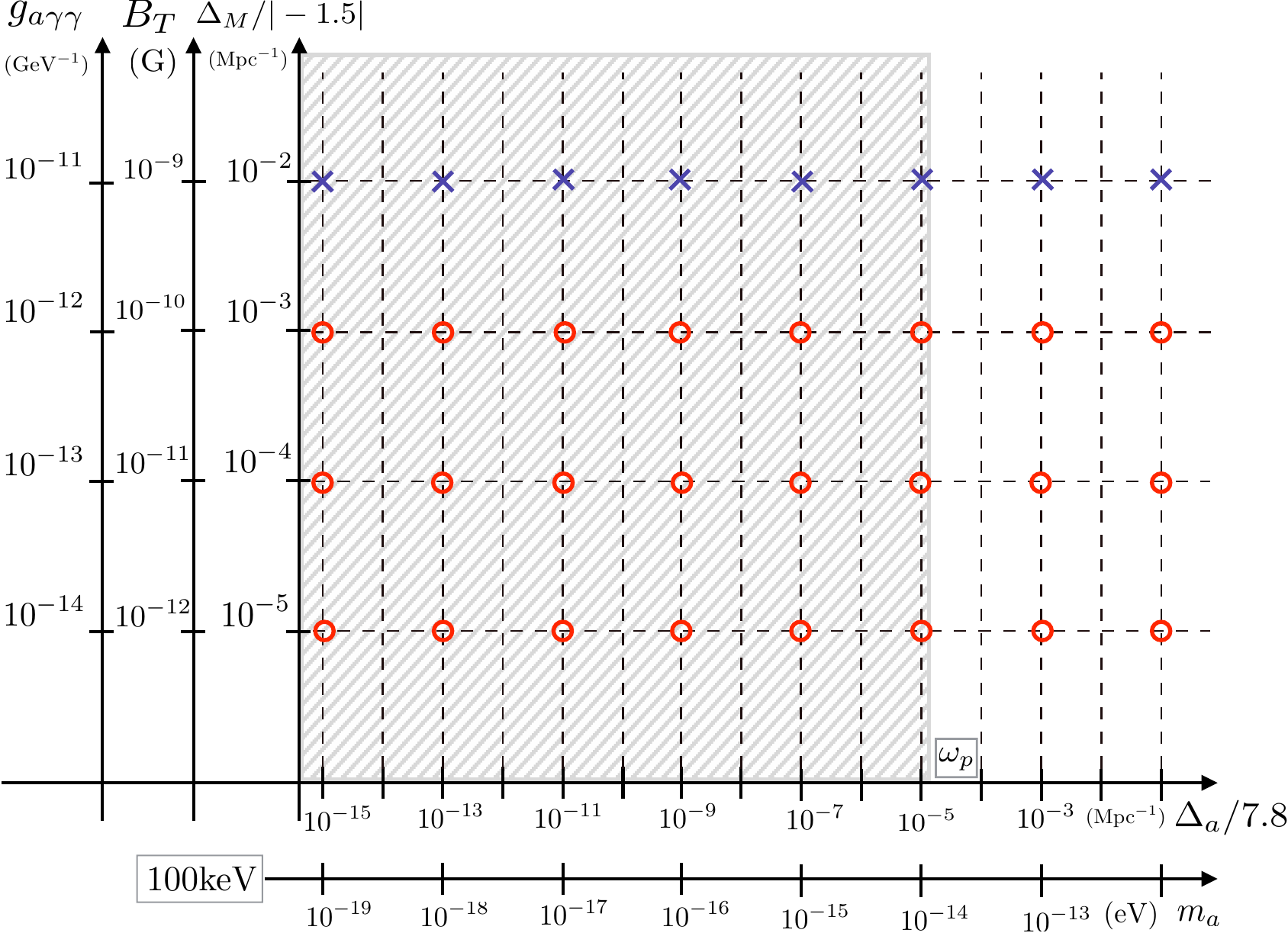}
    \caption{ Constraint for IGMF by 100\,keV photon for $|\alpha |\leq \pi \, \text{Mpc}^{-1}$.
    }
    \label{min}
    \end{center}
\end{figure}

We plotted the constraints on $(\Delta_a,\Delta_M)$ for the range $|\alpha |\leq \pi/180 \, \text{Mpc}^{-1}$ in Fig.~\ref{max}.
We took $\omega = 100$\,keV to give the constraints on the parameters.
The reason can be understood as follows.
For a parameter set $\Delta_{M} \ll \Delta_a $, there is no conversion.
Hence, $\Pi_C$ is small.
For a parameter set $\Delta_{a} \ll \Delta_M$, the photon-axion conversion can occur efficiently, but $\Pi_C$ is small.
This is because eigenvalues of mixing matrix go as $\lambda_+\sim |\lambda_-| \sim\Delta_M$, and no phase difference appears between two photon components \cite{a-p:2}.
Thus, the parameters satisfying
 $\Delta_a \sim \Delta_M \sim 1/N s \sim 10^{-4}~{\rm Mpc}^{-1}$
are most effective for generating the circular polarization.
Note that, once the condition $\Delta_a<\Delta_{plasma}$ is achieved, $\Pi_C$ does not depend on $\Delta_a$.
Hence, it is possible to give the strongest constraint when $\Delta_{plasma}\sim10^{-4}~{\rm Mpc}^{-1}$.
That is why we chose the energy of photon as 100\,keV.
The above argument also implies that the constraints are the same for $m_a \lesssim 10^{-14}$ eV.
The constraints in the helical domain model with the range $|\alpha | \leq \pi \, \text{Mpc}^{-1}$ can be seen in Fig.~\ref{min}.
We found that the constraints in the conventional model is also similar to this case.
We can see the constraints on the parameters are weak as is expected.

\subsection{Discussion}\label{4}

In \cite{polarization:9,polarization:11}, they focus on the magnetic field in a galaxy or a cluster of galaxies.
Although they mentioned the possibility that field strength, coherence length, and orientation can be rescaled, they fixed strength of galactic magnetic field to be several $\mu{\rm G}$ and gave a constraint on the coupling constant $g_{a\gamma\gamma}$.

On the other hand, we envisage that intergalactic magnetic fields, which have only the upper and the lower bound, cause the photon-axion conversion.
Moreover, we consider helical magnetic fields, provided that the origin of cosmological magnetic field is primordial.
The plausible value for coupling constant $g_{a\gamma\gamma}$ is also unknown, though there are suggestions from astrophysics.
Hence, we give a constraint on either coupling constant $g_{a\gamma\gamma}$ or $B_T$ by fixing one of them.

If we fix the coupling constant $g_{a\gamma\gamma}\simeq10^{-11}~\rm{GeV}^{-1}$ and use the photon with the energy $100\,\rm{keV}$, then the constraint on the parameter set $(\Delta_a,\Delta_M)$ can be interpreted as a constraint on the IGMF 
\begin{equation}
  B_T < 10^{-11}\,{\rm G},\ \ \text{for}\ m_a \lesssim10^{-14}~{\rm eV}\ .
\end{equation}
Recently, the more stringent constraint on the coupling constant $g_{a\gamma\gamma}$ is claimed~\cite{experiment5}
\begin{equation*}
  g_{a\gamma\gamma}\lesssim 5.3\times10^{-12}\,{\rm GeV}^{-1},\ {\text for}\ m_a \lesssim 4.4\times10^{-10}~\rm{eV}\ .
\end{equation*}
In this case, the axis of magnetic field is rescaled, so our result is modified as
\begin{equation}
  B_T < 10^{-10}\,{\rm G}, \ \ \text{for}\ m_a \lesssim 10^{-14}\,{\rm eV}\ .
\end{equation}

Next, we constrain the coupling constant $g_{a\gamma\gamma}$.
If we fix the strength of the magnetic field to be nG and use the photon with the energy $100\,\rm{keV}$, then we can provide a stringent constraint on the coupling constant
\begin{equation}
 g_{a\gamma\gamma} < 10^{-13} {\rm GeV}^{-1}, \ \ \text{for}\ m_a \lesssim 10^{-14}\,{\rm eV}\ .
\end{equation}

Due to the lack of knowledge about intergalactic magnetic fields, we do not know how to choose the range of $\alpha$, but as we have mentioned in \ref{astro_helical}, the significant change of magnetic configuration seem to prevent the growth of circular polarization in the early stage of evolution.
So when we choose $|\alpha|\leq \pi/180\,\rm{rad}\,\rm{Mpc^{-1}}$, $\omega=100\,{\rm keV}$, we can give the most stringent constraint for $\Delta_a$-$\Delta_M$ plane.

There are several studies on photon-axion conversion in helical model~\cite{helical:1,helical:2,helical:3}.
However, we want to stress again that our helical configuration is different from all of them and our configuration is more general. 
In this paper, we show that our model becomes similar to conventional model, when we make the range for $\alpha$ broad.
In addition, we show that when we make the range for $\alpha$ narrow, transient behavior of polarization reflect configuration difference and it is relevant to constraint for physical parameters.

\section{Conclusion}\label{conclusion}

We studied the magnetic field configuration dependence on the photon-axion conversion process.
Although most previous studies have been carried out in a simple model where each domain of a network of domains has a constant magnetic field, we studied a more general model where a network of domains are still assumed, but each domain has a helical magnetic field.
Recently, Wang and Lai \cite{helical:3} claimed that the photon-axion conversion in helical model behaves peculiarly.
As mentioned in the section \ref{sec3:configuration}, we find that our helical model gives much closer predictions to the conventional discontinuous magnetic field configuration model.  Our model takes $\alpha$ as a random variable, but Wang and Lai take $\varphi$ as a random variable.
Note that how to connect domains affects photon-axion conversion in helical model.

Moreover, we investigated the evolution of polarization during the photon-axion conversion process.
Remarkably, we analytically obtained asymptotic values of the variance of polarization in the conventional model.
We found that the asymptotic behavior does not depend on the configuration of magnetic fields.
While, when the helicity is small, we showed the oscillation appears in the early phase of evolution.
Moreover, the constraints on physical parameters obtained by the polarization observations depend on the magnetic field configuration.  
This is because transient behavior depends on the configuration.

We also presented phenomenological applications.
More precisely, taking into account that the degree of circular polarization from GRBs must be smaller than 1\%~\cite{experiment4}, we constrained the parameter $\Delta_{M} = -(1 / 2)g_{a\gamma\gamma}B_{T}$ so that the photon-axion conversion does not contradict with the observations.
The constraint on $\Delta_{M}$ can be translated to the constraint on $g_{a\gamma\gamma}$ or $B_{T}$ when we fix one of them to an appropriate value.
For the magnetic field configuration with the range $|\alpha | \leq \pi/180 \, \text{Mpc}^{-1}$, we obtained the stringent constraint on IGMF
\begin{equation*}
  B_T < 10^{-11}\rm{G},\ \ \text{for}\ m_a\lesssim10^{-14}~{\rm eV}\ , 
\end{equation*}
under the assumption $g_{a\gamma\gamma}= 10^{-11}\,\rm{GeV}^{-1}$.
Alternatively, by assuming the magnitude of magnetic field $|\bm{B}_{IGMF}| = 10^{-9}~\rm{G}$, we obtained the constraint on the coupling constant
\begin{equation*}
  g_{a\gamma\gamma}<10^{-13}\,\rm{GeV}^{-1},\ {\text for}\ m_a \lesssim 10^{-14}~\rm{eV}\ .
\end{equation*}
However, since the choice of the range of $\alpha$ affects the circular polarization of photons, we need to be careful.
Indeed, we have to know the magnetic field configuration in order to obtain the more precise constraint.

Admittedly, the models we have considered are still restrictive.
However, we believe the asymptotic behavior revealed in this paper is universal.

\acknowledgments
We would like to thank Kazunori Kohri for useful discussion.
This work was in part supported by JSPS Grant-in-Aid for JSPS Research
Fellow Grant Number 17J00568 (A.A.), JSPS KAKENHI Grant Numbers
17H02894, 17K18778, and MEXT KAKENHI Grant Numbers 15H05895, 17H06359
(J.S.).

\nocite{*}
\bibliographystyle{apsrev4-1}
\bibliography{0829}

\begin{thebibliography}{49}%
\makeatletter
\providecommand \@ifxundefined [1]{%
 \@ifx{#1\undefined}
}%
\providecommand \@ifnum [1]{%
 \ifnum #1\expandafter \@firstoftwo
 \else \expandafter \@secondoftwo
 \fi
}%
\providecommand \@ifx [1]{%
 \ifx #1\expandafter \@firstoftwo
 \else \expandafter \@secondoftwo
 \fi
}%
\providecommand \natexlab [1]{#1}%
\providecommand \enquote  [1]{``#1''}%
\providecommand \bibnamefont  [1]{#1}%
\providecommand \bibfnamefont [1]{#1}%
\providecommand \citenamefont [1]{#1}%
\providecommand \href@noop [0]{\@secondoftwo}%
\providecommand \href [0]{\begingroup \@sanitize@url \@href}%
\providecommand \@href[1]{\@@startlink{#1}\@@href}%
\providecommand \@@href[1]{\endgroup#1\@@endlink}%
\providecommand \@sanitize@url [0]{\catcode `\\12\catcode `\$12\catcode
  `\&12\catcode `\#12\catcode `\^12\catcode `\_12\catcode `\%12\relax}%
\providecommand \@@startlink[1]{}%
\providecommand \@@endlink[0]{}%
\providecommand \url  [0]{\begingroup\@sanitize@url \@url }%
\providecommand \@url [1]{\endgroup\@href {#1}{\urlprefix }}%
\providecommand \urlprefix  [0]{URL }%
\providecommand \Eprint [0]{\href }%
\providecommand \doibase [0]{http://dx.doi.org/}%
\providecommand \selectlanguage [0]{\@gobble}%
\providecommand \bibinfo  [0]{\@secondoftwo}%
\providecommand \bibfield  [0]{\@secondoftwo}%
\providecommand \translation [1]{[#1]}%
\providecommand \BibitemOpen [0]{}%
\providecommand \bibitemStop [0]{}%
\providecommand \bibitemNoStop [0]{.\EOS\space}%
\providecommand \EOS [0]{\spacefactor3000\relax}%
\providecommand \BibitemShut  [1]{\csname bibitem#1\endcsname}%
\let\auto@bib@innerbib\@empty
\bibitem [{\citenamefont {Peccei}\ and\ \citenamefont
  {Quinn}(1977{\natexlab{a}})}]{QCD1}%
  \BibitemOpen
  \bibfield  {author} {\bibinfo {author} {\bibfnamefont {R.~D.}\ \bibnamefont
  {Peccei}}\ and\ \bibinfo {author} {\bibfnamefont {H.~R.}\ \bibnamefont
  {Quinn}},\ }\href {\doibase 10.1103/PhysRevLett.38.1440} {\bibfield
  {journal} {\bibinfo  {journal} {Phys. Rev. Lett.}\ }\textbf {\bibinfo
  {volume} {38}},\ \bibinfo {pages} {1440} (\bibinfo {year}
  {1977}{\natexlab{a}})}\BibitemShut {NoStop}%
\bibitem [{\citenamefont {Peccei}\ and\ \citenamefont
  {Quinn}(1977{\natexlab{b}})}]{QCD2}%
  \BibitemOpen
  \bibfield  {author} {\bibinfo {author} {\bibfnamefont {R.~D.}\ \bibnamefont
  {Peccei}}\ and\ \bibinfo {author} {\bibfnamefont {H.~R.}\ \bibnamefont
  {Quinn}},\ }\href {\doibase 10.1103/PhysRevD.16.1791} {\bibfield  {journal}
  {\bibinfo  {journal} {Phys. Rev. D}\ }\textbf {\bibinfo {volume} {16}},\
  \bibinfo {pages} {1791} (\bibinfo {year} {1977}{\natexlab{b}})}\BibitemShut
  {NoStop}%
\bibitem [{\citenamefont {Weinberg}(1978)}]{QCD3}%
  \BibitemOpen
  \bibfield  {author} {\bibinfo {author} {\bibfnamefont {S.}~\bibnamefont
  {Weinberg}},\ }\href {\doibase 10.1103/PhysRevLett.40.223} {\bibfield
  {journal} {\bibinfo  {journal} {Phys. Rev. Lett.}\ }\textbf {\bibinfo
  {volume} {40}},\ \bibinfo {pages} {223} (\bibinfo {year} {1978})}\BibitemShut
  {NoStop}%
\bibitem [{\citenamefont {Wilczek}(1978)}]{QCD4}%
  \BibitemOpen
  \bibfield  {author} {\bibinfo {author} {\bibfnamefont {F.}~\bibnamefont
  {Wilczek}},\ }\href {\doibase 10.1103/PhysRevLett.40.279} {\bibfield
  {journal} {\bibinfo  {journal} {Phys. Rev. Lett.}\ }\textbf {\bibinfo
  {volume} {40}},\ \bibinfo {pages} {279} (\bibinfo {year} {1978})}\BibitemShut
  {NoStop}%
\bibitem [{\citenamefont {Svrcek}\ and\ \citenamefont
  {Witten}(2006)}]{String1}%
  \BibitemOpen
  \bibfield  {author} {\bibinfo {author} {\bibfnamefont {P.}~\bibnamefont
  {Svrcek}}\ and\ \bibinfo {author} {\bibfnamefont {E.}~\bibnamefont
  {Witten}},\ }\href {\doibase 10.1088/1126-6708/2006/06/051} {\bibfield
  {journal} {\bibinfo  {journal} {J. High Energy Phys.}\ }\textbf {\bibinfo
  {volume} {2006}},\ \bibinfo {pages} {051} (\bibinfo {year} {2006})},\ \Eprint
  {http://arxiv.org/abs/hep-th/0605206} {arXiv:hep-th/0605206 [hep-th]}
  \BibitemShut {NoStop}%
\bibitem [{\citenamefont {Arvanitaki}\ \emph {et~al.}(2010)\citenamefont
  {Arvanitaki}, \citenamefont {Dimopoulos}, \citenamefont {Dubovsky},
  \citenamefont {Kaloper} \emph {et~al.}}]{String2}%
  \BibitemOpen
  \bibfield  {author} {\bibinfo {author} {\bibfnamefont {A.}~\bibnamefont
  {Arvanitaki}}, \bibinfo {author} {\bibfnamefont {S.}~\bibnamefont
  {Dimopoulos}}, \bibinfo {author} {\bibfnamefont {S.}~\bibnamefont
  {Dubovsky}}, \bibinfo {author} {\bibfnamefont {N.}~\bibnamefont {Kaloper}},
  \emph {et~al.},\ }\href {\doibase 10.1103/PhysRevD.81.123530} {\bibfield
  {journal} {\bibinfo  {journal} {Phys. Rev. D}\ }\textbf {\bibinfo {volume}
  {81}},\ \bibinfo {pages} {123530} (\bibinfo {year} {2010})},\ \Eprint
  {http://arxiv.org/abs/0905.4720} {arXiv:0905.4720 [hep-th]} \BibitemShut
  {NoStop}%
\bibitem [{\citenamefont {Durrer}\ and\ \citenamefont
  {Neronov}(2013)}]{Magnetic_Field1}%
  \BibitemOpen
  \bibfield  {author} {\bibinfo {author} {\bibfnamefont {R.}~\bibnamefont
  {Durrer}}\ and\ \bibinfo {author} {\bibfnamefont {A.}~\bibnamefont
  {Neronov}},\ }\href {\doibase 10.1007/s00159-013-0062-7} {\bibfield
  {journal} {\bibinfo  {journal} {The Astronomy and Astrophysics Review}\
  }\textbf {\bibinfo {volume} {21}},\ \bibinfo {pages} {62} (\bibinfo {year}
  {2013})},\ \Eprint {http://arxiv.org/abs/1303.7121} {arXiv:1303.7121
  [astro-ph.CO]} \BibitemShut {NoStop}%
\bibitem [{\citenamefont {Subramanian}(2016)}]{Magnetic_Field2}%
  \BibitemOpen
  \bibfield  {author} {\bibinfo {author} {\bibfnamefont {K.}~\bibnamefont
  {Subramanian}},\ }\href {\doibase 10.1088/0034-4885/79/7/076901} {\bibfield
  {journal} {\bibinfo  {journal} {Reports on Progress in Physics}\ }\textbf
  {\bibinfo {volume} {79}},\ \bibinfo {pages} {076901} (\bibinfo {year}
  {2016})},\ \Eprint {http://arxiv.org/abs/1504.02311} {arXiv:1504.02311
  [astro-ph.CO]} \BibitemShut {NoStop}%
\bibitem [{\citenamefont {Neronov}\ and\ \citenamefont
  {Vovk}(2010)}]{experiment3}%
  \BibitemOpen
  \bibfield  {author} {\bibinfo {author} {\bibfnamefont {A.}~\bibnamefont
  {Neronov}}\ and\ \bibinfo {author} {\bibfnamefont {I.}~\bibnamefont {Vovk}},\
  }\href {\doibase 10.1126/science.1184192} {\bibfield  {journal} {\bibinfo
  {journal} {Science}\ }\textbf {\bibinfo {volume} {328}},\ \bibinfo {pages}
  {73} (\bibinfo {year} {2010})},\ \Eprint {http://arxiv.org/abs/1006.3504}
  {arXiv:1006.3504 [astro-ph]} \BibitemShut {NoStop}%
\bibitem [{\citenamefont {Maiani}\ \emph {et~al.}(1986)\citenamefont {Maiani},
  \citenamefont {Petronzio},\ and\ \citenamefont {Zavattini}}]{a-p:1}%
  \BibitemOpen
  \bibfield  {author} {\bibinfo {author} {\bibfnamefont {L.}~\bibnamefont
  {Maiani}}, \bibinfo {author} {\bibfnamefont {R.}~\bibnamefont {Petronzio}}, \
  and\ \bibinfo {author} {\bibfnamefont {E.}~\bibnamefont {Zavattini}},\ }\href
  {\doibase 10.1016/0370-2693(86)90869-5} {\bibfield  {journal} {\bibinfo
  {journal} {Phys. Lett. B}\ }\textbf {\bibinfo {volume} {175}},\ \bibinfo
  {pages} {359} (\bibinfo {year} {1986})}\BibitemShut {NoStop}%
\bibitem [{\citenamefont {Raffelt}\ and\ \citenamefont
  {Stodolsky}(1988)}]{a-p:2}%
  \BibitemOpen
  \bibfield  {author} {\bibinfo {author} {\bibfnamefont {G.}~\bibnamefont
  {Raffelt}}\ and\ \bibinfo {author} {\bibfnamefont {L.}~\bibnamefont
  {Stodolsky}},\ }\href {\doibase 10.1103/PhysRevD.37.1237} {\bibfield
  {journal} {\bibinfo  {journal} {Phys. Rev. D}\ }\textbf {\bibinfo {volume}
  {37}},\ \bibinfo {pages} {1237} (\bibinfo {year} {1988})}\BibitemShut
  {NoStop}%
\bibitem [{\citenamefont {Cs\'aki}\ \emph
  {et~al.}(2002{\natexlab{a}})\citenamefont {Cs\'aki}, \citenamefont
  {Kaloper},\ and\ \citenamefont {Terning}}]{Ia:1}%
  \BibitemOpen
  \bibfield  {author} {\bibinfo {author} {\bibfnamefont {C.}~\bibnamefont
  {Cs\'aki}}, \bibinfo {author} {\bibfnamefont {N.}~\bibnamefont {Kaloper}}, \
  and\ \bibinfo {author} {\bibfnamefont {J.}~\bibnamefont {Terning}},\ }\href
  {\doibase 10.1016/S0370-2693(02)01765-3} {\bibfield  {journal} {\bibinfo
  {journal} {Phys. Lett. B}\ }\textbf {\bibinfo {volume} {535}},\ \bibinfo
  {pages} {33 } (\bibinfo {year} {2002}{\natexlab{a}})},\ \Eprint
  {http://arxiv.org/abs/hep-ph/0112212} {arXiv:hep-ph/0112212 [hep-ph]}
  \BibitemShut {NoStop}%
\bibitem [{\citenamefont {Cs\'aki}\ \emph
  {et~al.}(2002{\natexlab{b}})\citenamefont {Cs\'aki}, \citenamefont
  {Kaloper},\ and\ \citenamefont {Terning}}]{Ia:2}%
  \BibitemOpen
  \bibfield  {author} {\bibinfo {author} {\bibfnamefont {C.}~\bibnamefont
  {Cs\'aki}}, \bibinfo {author} {\bibfnamefont {N.}~\bibnamefont {Kaloper}}, \
  and\ \bibinfo {author} {\bibfnamefont {J.}~\bibnamefont {Terning}},\ }\href
  {\doibase 10.1103/PhysRevLett.88.161302} {\bibfield  {journal} {\bibinfo
  {journal} {Phys. Rev. Lett.}\ }\textbf {\bibinfo {volume} {88}},\ \bibinfo
  {pages} {16} (\bibinfo {year} {2002}{\natexlab{b}})},\ \Eprint
  {http://arxiv.org/abs/hep-ph/0112212} {arXiv:hep-ph/0112212} \BibitemShut
  {NoStop}%
\bibitem [{\citenamefont {Deffayet}\ \emph {et~al.}(2002)\citenamefont
  {Deffayet}, \citenamefont {Harari}, \citenamefont {Uzan},\ and\ \citenamefont
  {Zaldarriaga}}]{Ia:3}%
  \BibitemOpen
  \bibfield  {author} {\bibinfo {author} {\bibfnamefont {C.}~\bibnamefont
  {Deffayet}}, \bibinfo {author} {\bibfnamefont {D.}~\bibnamefont {Harari}},
  \bibinfo {author} {\bibfnamefont {J.-P.}\ \bibnamefont {Uzan}}, \ and\
  \bibinfo {author} {\bibfnamefont {M.}~\bibnamefont {Zaldarriaga}},\ }\href
  {\doibase 10.1103/PhysRevD.66.043517} {\bibfield  {journal} {\bibinfo
  {journal} {Phys. Rev. D}\ }\textbf {\bibinfo {volume} {66}},\ \bibinfo
  {pages} {043517} (\bibinfo {year} {2002})},\ \Eprint
  {http://arxiv.org/abs/hep-ph/0112118} {arXiv:hep-ph/0112118} \BibitemShut
  {NoStop}%
\bibitem [{\citenamefont {Grossman}\ \emph {et~al.}(2002)\citenamefont
  {Grossman}, \citenamefont {Roy},\ and\ \citenamefont {Zupan}}]{Ia:4}%
  \BibitemOpen
  \bibfield  {author} {\bibinfo {author} {\bibfnamefont {Y.}~\bibnamefont
  {Grossman}}, \bibinfo {author} {\bibfnamefont {S.}~\bibnamefont {Roy}}, \
  and\ \bibinfo {author} {\bibfnamefont {J.}~\bibnamefont {Zupan}},\ }\href
  {\doibase 10.1016/S0370-2693(02)02448-6} {\bibfield  {journal} {\bibinfo
  {journal} {Phys. Lett. B}\ }\textbf {\bibinfo {volume} {543}},\ \bibinfo
  {pages} {23 } (\bibinfo {year} {2002})},\ \Eprint
  {http://arxiv.org/abs/hep-ph/0204216} {arXiv:hep-ph/0204216 [hep-ph]}
  \BibitemShut {NoStop}%
\bibitem [{\citenamefont {Christensson}\ and\ \citenamefont
  {Fairbairn}(2003)}]{Ia:5}%
  \BibitemOpen
  \bibfield  {author} {\bibinfo {author} {\bibfnamefont {M.}~\bibnamefont
  {Christensson}}\ and\ \bibinfo {author} {\bibfnamefont {M.}~\bibnamefont
  {Fairbairn}},\ }\href {\doibase 10.1016/S0370-2693(03)00641-5} {\bibfield
  {journal} {\bibinfo  {journal} {Phys. Lett. B}\ }\textbf {\bibinfo {volume}
  {565}},\ \bibinfo {pages} {10} (\bibinfo {year} {2003})},\ \Eprint
  {http://arxiv.org/abs/astro-ph/0207525} {arXiv:astro-ph/0207525 [astro-ph]}
  \BibitemShut {NoStop}%
\bibitem [{\citenamefont {De~Angelis}\ \emph {et~al.}(2007)\citenamefont
  {De~Angelis}, \citenamefont {Roncadelli},\ and\ \citenamefont
  {Mansutti}}]{transparency:1}%
  \BibitemOpen
  \bibfield  {author} {\bibinfo {author} {\bibfnamefont {A.}~\bibnamefont
  {De~Angelis}}, \bibinfo {author} {\bibfnamefont {M.}~\bibnamefont
  {Roncadelli}}, \ and\ \bibinfo {author} {\bibfnamefont {O.}~\bibnamefont
  {Mansutti}},\ }\href {\doibase 10.1103/PhysRevD.76.121301} {\bibfield
  {journal} {\bibinfo  {journal} {Phys. Rev. D}\ }\textbf {\bibinfo {volume}
  {76}},\ \bibinfo {pages} {121301} (\bibinfo {year} {2007})},\ \Eprint
  {http://arxiv.org/abs/0707.4312} {arXiv:0707.4312 [astro-ph]} \BibitemShut
  {NoStop}%
\bibitem [{\citenamefont {Simet}\ \emph {et~al.}(2008)\citenamefont {Simet},
  \citenamefont {Hooper},\ and\ \citenamefont {Serpico}}]{transparency:2}%
  \BibitemOpen
  \bibfield  {author} {\bibinfo {author} {\bibfnamefont {M.}~\bibnamefont
  {Simet}}, \bibinfo {author} {\bibfnamefont {D.}~\bibnamefont {Hooper}}, \
  and\ \bibinfo {author} {\bibfnamefont {P.~D.}\ \bibnamefont {Serpico}},\
  }\href {\doibase 10.1103/PhysRevD.77.063001} {\bibfield  {journal} {\bibinfo
  {journal} {Phys. Rev. D}\ }\textbf {\bibinfo {volume} {77}},\ \bibinfo
  {pages} {063001} (\bibinfo {year} {2008})},\ \Eprint
  {http://arxiv.org/abs/0712.2825} {arXiv:0712.2825 [astro-ph]} \BibitemShut
  {NoStop}%
\bibitem [{\citenamefont {S\'anchez-Conde}\ \emph {et~al.}(2009)\citenamefont
  {S\'anchez-Conde}, \citenamefont {Paneque}, \citenamefont {Bloom},
  \citenamefont {Prada},\ and\ \citenamefont
  {Dom\'{\i}nguez}}]{transparency:3}%
  \BibitemOpen
  \bibfield  {author} {\bibinfo {author} {\bibfnamefont {M.~A.}\ \bibnamefont
  {S\'anchez-Conde}}, \bibinfo {author} {\bibfnamefont {D.}~\bibnamefont
  {Paneque}}, \bibinfo {author} {\bibfnamefont {E.}~\bibnamefont {Bloom}},
  \bibinfo {author} {\bibfnamefont {F.}~\bibnamefont {Prada}}, \ and\ \bibinfo
  {author} {\bibfnamefont {A.}~\bibnamefont {Dom\'{\i}nguez}},\ }\href
  {\doibase 10.1103/PhysRevD.79.123511} {\bibfield  {journal} {\bibinfo
  {journal} {Phys. Rev. D}\ }\textbf {\bibinfo {volume} {79}},\ \bibinfo
  {pages} {123511} (\bibinfo {year} {2009})},\ \Eprint
  {http://arxiv.org/abs/0905.3270} {arXiv:0905.3270 [astro-ph.CO]} \BibitemShut
  {NoStop}%
\bibitem [{\citenamefont {Mirizzi}\ and\ \citenamefont
  {Montanino}(2009)}]{transparency:4}%
  \BibitemOpen
  \bibfield  {author} {\bibinfo {author} {\bibfnamefont {A.}~\bibnamefont
  {Mirizzi}}\ and\ \bibinfo {author} {\bibfnamefont {D.}~\bibnamefont
  {Montanino}},\ }\href {\doibase 10.1088/1475-7516/2009/12/004} {\bibfield
  {journal} {\bibinfo  {journal} {J. Cosmol. Astropart. Phys}\ }\textbf
  {\bibinfo {volume} {2009}},\ \bibinfo {pages} {004} (\bibinfo {year}
  {2009})},\ \Eprint {http://arxiv.org/abs/0911.0015} {arXiv:0911.0015
  [astro-ph.HE]} \BibitemShut {NoStop}%
\bibitem [{\citenamefont {De~Angelis}\ \emph {et~al.}(2011)\citenamefont
  {De~Angelis}, \citenamefont {Galanti},\ and\ \citenamefont
  {Roncadelli}}]{transparency:5}%
  \BibitemOpen
  \bibfield  {author} {\bibinfo {author} {\bibfnamefont {A.}~\bibnamefont
  {De~Angelis}}, \bibinfo {author} {\bibfnamefont {G.}~\bibnamefont {Galanti}},
  \ and\ \bibinfo {author} {\bibfnamefont {M.}~\bibnamefont {Roncadelli}},\
  }\href {\doibase 10.1103/PhysRevD.84.105030} {\bibfield  {journal} {\bibinfo
  {journal} {Phys. Rev.D}\ }\textbf {\bibinfo {volume} {84}},\ \bibinfo {pages}
  {105030} (\bibinfo {year} {2011})},\ \Eprint {http://arxiv.org/abs/1106.1132}
  {arXiv:1106.1132 [astro-ph.HE]} \BibitemShut {NoStop}%
\bibitem [{\citenamefont {Horns}\ \emph
  {et~al.}(2012{\natexlab{a}})\citenamefont {Horns}, \citenamefont {Maccione},
  \citenamefont {Meyer}, \citenamefont {Mirizzi} \emph
  {et~al.}}]{transparency:6}%
  \BibitemOpen
  \bibfield  {author} {\bibinfo {author} {\bibfnamefont {D.}~\bibnamefont
  {Horns}}, \bibinfo {author} {\bibfnamefont {L.}~\bibnamefont {Maccione}},
  \bibinfo {author} {\bibfnamefont {M.}~\bibnamefont {Meyer}}, \bibinfo
  {author} {\bibfnamefont {A.}~\bibnamefont {Mirizzi}},  \emph {et~al.},\
  }\href {\doibase 10.1103/PhysRevD.86.075024} {\bibfield  {journal} {\bibinfo
  {journal} {Phys. Rev. D}\ }\textbf {\bibinfo {volume} {86}},\ \bibinfo
  {pages} {075024} (\bibinfo {year} {2012}{\natexlab{a}})},\ \Eprint
  {http://arxiv.org/abs/1207.0776} {arXiv:1207.0776 [astro-ph.HE]} \BibitemShut
  {NoStop}%
\bibitem [{\citenamefont {De~Angelis}\ \emph {et~al.}(2013)\citenamefont
  {De~Angelis}, \citenamefont {Galanti},\ and\ \citenamefont
  {Roncadelli}}]{transparency:7}%
  \BibitemOpen
  \bibfield  {author} {\bibinfo {author} {\bibfnamefont {A.}~\bibnamefont
  {De~Angelis}}, \bibinfo {author} {\bibfnamefont {G.}~\bibnamefont {Galanti}},
  \ and\ \bibinfo {author} {\bibfnamefont {M.}~\bibnamefont {Roncadelli}},\
  }\href {\doibase 10.1103/PhysRevD.87.109903} {\bibfield  {journal} {\bibinfo
  {journal} {Phys. Rev. D}\ }\textbf {\bibinfo {volume} {87}},\ \bibinfo
  {pages} {109903} (\bibinfo {year} {2013})},\ \Eprint
  {http://arxiv.org/abs/1106.1132} {arXiv:1106.1132 [astro-ph.HE]} \BibitemShut
  {NoStop}%
\bibitem [{\citenamefont {Abramowski}\ \emph {et~al.}(2013)\citenamefont
  {Abramowski} \emph {et~al.}}]{transparency:8}%
  \BibitemOpen
  \bibfield  {author} {\bibinfo {author} {\bibfnamefont {A.}~\bibnamefont
  {Abramowski}} \emph {et~al.},\ }\href {\doibase 10.1103/PhysRevD.88.102003}
  {\bibfield  {journal} {\bibinfo  {journal} {Phys. Rev. D}\ }\textbf {\bibinfo
  {volume} {88}},\ \bibinfo {pages} {102003} (\bibinfo {year} {2013})},\
  \Eprint {http://arxiv.org/abs/1311.3148} {arXiv:1311.3148 [astro-ph.HE]}
  \BibitemShut {NoStop}%
\bibitem [{\citenamefont {Meyer}\ \emph {et~al.}(2014)\citenamefont {Meyer},
  \citenamefont {Montanino},\ and\ \citenamefont {Conrad}}]{transparency:9}%
  \BibitemOpen
  \bibfield  {author} {\bibinfo {author} {\bibfnamefont {M.}~\bibnamefont
  {Meyer}}, \bibinfo {author} {\bibfnamefont {D.}~\bibnamefont {Montanino}}, \
  and\ \bibinfo {author} {\bibfnamefont {J.}~\bibnamefont {Conrad}},\ }\href
  {\doibase 10.1088/1475-7516/2014/09/003} {\bibfield  {journal} {\bibinfo
  {journal} {J. Cosmol. Astropart. Phys}\ }\textbf {\bibinfo {volume} {2014}},\
  \bibinfo {pages} {003} (\bibinfo {year} {2014})},\ \Eprint
  {http://arxiv.org/abs/1406.5972} {arXiv:1406.5972 [astro-ph.HE]} \BibitemShut
  {NoStop}%
\bibitem [{\citenamefont {Wouters}\ and\ \citenamefont
  {Brun}(2014)}]{transparency:10}%
  \BibitemOpen
  \bibfield  {author} {\bibinfo {author} {\bibfnamefont {D.}~\bibnamefont
  {Wouters}}\ and\ \bibinfo {author} {\bibfnamefont {P.}~\bibnamefont {Brun}},\
  }\href {\doibase 10.1088/1475-7516/2014/01/016} {\bibfield  {journal}
  {\bibinfo  {journal} {J. Cosmol. Astropart. Phys}\ }\textbf {\bibinfo
  {volume} {2014}},\ \bibinfo {pages} {016} (\bibinfo {year} {2014})},\ \Eprint
  {http://arxiv.org/abs/1309.6752} {arXiv:1309.6752 [astro-ph.HE]} \BibitemShut
  {NoStop}%
\bibitem [{\citenamefont {Kuster}\ \emph {et~al.}(2007)\citenamefont {Kuster},
  \citenamefont {Raffelt},\ and\ \citenamefont {Beltr{\'a}n}}]{others:1}%
  \BibitemOpen
  \bibfield  {author} {\bibinfo {author} {\bibfnamefont {M.}~\bibnamefont
  {Kuster}}, \bibinfo {author} {\bibfnamefont {G.}~\bibnamefont {Raffelt}}, \
  and\ \bibinfo {author} {\bibfnamefont {B.}~\bibnamefont {Beltr{\'a}n}},\
  }\href@noop {} {\emph {\bibinfo {title} {{Axions: Theory, cosmology, and
  experimental searches}}}},\ Vol.\ \bibinfo {volume} {741}\ (\bibinfo
  {publisher} {Springer},\ \bibinfo {year} {2007})\BibitemShut {NoStop}%
\bibitem [{\citenamefont {Ganguly}\ \emph {et~al.}(2009)\citenamefont
  {Ganguly}, \citenamefont {Jain},\ and\ \citenamefont {Mandal}}]{others:2}%
  \BibitemOpen
  \bibfield  {author} {\bibinfo {author} {\bibfnamefont {A.~K.}\ \bibnamefont
  {Ganguly}}, \bibinfo {author} {\bibfnamefont {P.}~\bibnamefont {Jain}}, \
  and\ \bibinfo {author} {\bibfnamefont {S.}~\bibnamefont {Mandal}},\ }\href
  {\doibase 10.1103/PhysRevD.79.115014} {\bibfield  {journal} {\bibinfo
  {journal} {Phys. Rev. D}\ }\textbf {\bibinfo {volume} {79}},\ \bibinfo
  {pages} {115014} (\bibinfo {year} {2009})},\ \Eprint
  {http://arxiv.org/abs/0810.4380} {arXiv:0810.4380 [hep-ph]} \BibitemShut
  {NoStop}%
\bibitem [{\citenamefont {Burrage}\ \emph {et~al.}(2009)\citenamefont
  {Burrage}, \citenamefont {Davis},\ and\ \citenamefont {Shaw}}]{others:3}%
  \BibitemOpen
  \bibfield  {author} {\bibinfo {author} {\bibfnamefont {C.}~\bibnamefont
  {Burrage}}, \bibinfo {author} {\bibfnamefont {A.-C.}\ \bibnamefont {Davis}},
  \ and\ \bibinfo {author} {\bibfnamefont {D.~J.}\ \bibnamefont {Shaw}},\
  }\href {\doibase 10.1103/PhysRevLett.102.201101} {\bibfield  {journal}
  {\bibinfo  {journal} {Phys. Rev. Lett.}\ }\textbf {\bibinfo {volume} {102}},\
  \bibinfo {pages} {201101} (\bibinfo {year} {2009})},\ \Eprint
  {http://arxiv.org/abs/0902.2320} {arXiv:0902.2320 [astro-ph.CO]} \BibitemShut
  {NoStop}%
\bibitem [{\citenamefont {Fairbairn}\ \emph {et~al.}(2011)\citenamefont
  {Fairbairn}, \citenamefont {Rashba},\ and\ \citenamefont
  {Troitsky}}]{others:4}%
  \BibitemOpen
  \bibfield  {author} {\bibinfo {author} {\bibfnamefont {M.}~\bibnamefont
  {Fairbairn}}, \bibinfo {author} {\bibfnamefont {T.}~\bibnamefont {Rashba}}, \
  and\ \bibinfo {author} {\bibfnamefont {S.~V.}\ \bibnamefont {Troitsky}},\
  }\href {\doibase 10.1103/PhysRevD.84.125019} {\bibfield  {journal} {\bibinfo
  {journal} {Phys. Rev. D}\ }\textbf {\bibinfo {volume} {84}},\ \bibinfo
  {pages} {125019} (\bibinfo {year} {2011})},\ \Eprint
  {http://arxiv.org/abs/0901.4085} {arXiv:0901.4085 [astro-ph.HE]} \BibitemShut
  {NoStop}%
\bibitem [{\citenamefont {Das}\ \emph {et~al.}(2005)\citenamefont {Das},
  \citenamefont {Jain}, \citenamefont {Ralston} \emph {et~al.}}]{helical:1}%
  \BibitemOpen
  \bibfield  {author} {\bibinfo {author} {\bibfnamefont {S.}~\bibnamefont
  {Das}}, \bibinfo {author} {\bibfnamefont {P.}~\bibnamefont {Jain}}, \bibinfo
  {author} {\bibfnamefont {J.~P.}\ \bibnamefont {Ralston}},  \emph {et~al.},\
  }\href {\doibase 10.1088/1475-7516/2005/06/002} {\bibfield  {journal}
  {\bibinfo  {journal} {J. Cosmol. Astropart. Phys}\ }\textbf {\bibinfo
  {volume} {2005}},\ \bibinfo {pages} {002} (\bibinfo {year} {2005})},\ \Eprint
  {http://arxiv.org/abs/hep-ph/0408198} {arXiv:hep-ph/0408198 [hep-ph]}
  \BibitemShut {NoStop}%
\bibitem [{\citenamefont {Das}\ \emph {et~al.}(2008)\citenamefont {Das},
  \citenamefont {Jain}, \citenamefont {Ralston} \emph {et~al.}}]{helical:2}%
  \BibitemOpen
  \bibfield  {author} {\bibinfo {author} {\bibfnamefont {S.}~\bibnamefont
  {Das}}, \bibinfo {author} {\bibfnamefont {P.}~\bibnamefont {Jain}}, \bibinfo
  {author} {\bibfnamefont {J.~P.}\ \bibnamefont {Ralston}},  \emph {et~al.},\
  }\href {\doibase 10.1007/s12043-008-0060-x} {\bibfield  {journal} {\bibinfo
  {journal} {Pramana}\ }\textbf {\bibinfo {volume} {70}},\ \bibinfo {pages}
  {439} (\bibinfo {year} {2008})},\ \Eprint
  {http://arxiv.org/abs/hep-ph/0410006} {arXiv:hep-ph/0410006 [hep-ph]}
  \BibitemShut {NoStop}%
\bibitem [{\citenamefont {Wang}\ and\ \citenamefont {Lai}(2016)}]{helical:3}%
  \BibitemOpen
  \bibfield  {author} {\bibinfo {author} {\bibfnamefont {C.}~\bibnamefont
  {Wang}}\ and\ \bibinfo {author} {\bibfnamefont {D.}~\bibnamefont {Lai}},\
  }\href {\doibase 10.1088/1475-7516/2016/06/006} {\bibfield  {journal}
  {\bibinfo  {journal} {J. Cosmol. Astropart. Phys}\ }\textbf {\bibinfo
  {volume} {2016}},\ \bibinfo {pages} {006} (\bibinfo {year} {2016})},\ \Eprint
  {http://arxiv.org/abs/1511.03380} {arXiv:1511.03380 [astro-ph.HE]}
  \BibitemShut {NoStop}%
\bibitem [{\citenamefont {Wiersema}\ \emph {et~al.}(2014)\citenamefont
  {Wiersema} \emph {et~al.}}]{experiment4}%
  \BibitemOpen
  \bibfield  {author} {\bibinfo {author} {\bibfnamefont {K.}~\bibnamefont
  {Wiersema}} \emph {et~al.},\ }\href {\doibase 10.1038/nature13237} {\bibfield
   {journal} {\bibinfo  {journal} {Nature}\ }\textbf {\bibinfo {volume}
  {509}},\ \bibinfo {pages} {201} (\bibinfo {year} {2014})},\ \Eprint
  {http://arxiv.org/abs/1410.0489} {arXiv:1410.0489 [astro-ph.HE]} \BibitemShut
  {NoStop}%
\bibitem [{\citenamefont {Bassan}\ \emph {et~al.}(2010)\citenamefont {Bassan},
  \citenamefont {Mirizzi},\ and\ \citenamefont {Roncadelli}}]{polarization:4}%
  \BibitemOpen
  \bibfield  {author} {\bibinfo {author} {\bibfnamefont {N.}~\bibnamefont
  {Bassan}}, \bibinfo {author} {\bibfnamefont {A.}~\bibnamefont {Mirizzi}}, \
  and\ \bibinfo {author} {\bibfnamefont {M.}~\bibnamefont {Roncadelli}},\
  }\href {\doibase 10.1088/1475-7516/2010/05/010} {\bibfield  {journal}
  {\bibinfo  {journal} {J. Cosmol. Astropart. Phys}\ }\textbf {\bibinfo
  {volume} {2010}},\ \bibinfo {pages} {010} (\bibinfo {year} {2010})},\ \Eprint
  {http://arxiv.org/abs/1001.5267} {arXiv:1001.5267 [astro-ph.HE]} \BibitemShut
  {NoStop}%
\bibitem [{\citenamefont {Harari}\ and\ \citenamefont
  {Sikivie}(1992)}]{polarization:1}%
  \BibitemOpen
  \bibfield  {author} {\bibinfo {author} {\bibfnamefont {D.}~\bibnamefont
  {Harari}}\ and\ \bibinfo {author} {\bibfnamefont {P.}~\bibnamefont
  {Sikivie}},\ }\href {\doibase 10.1016/0370-2693(92)91363-E} {\bibfield
  {journal} {\bibinfo  {journal} {Phys. Lett. B}\ }\textbf {\bibinfo {volume}
  {289}},\ \bibinfo {pages} {67 } (\bibinfo {year} {1992})}\BibitemShut
  {NoStop}%
\bibitem [{\citenamefont {Jain}\ \emph {et~al.}(2002)\citenamefont {Jain},
  \citenamefont {Panda},\ and\ \citenamefont {Sarala}}]{polarization:2}%
  \BibitemOpen
  \bibfield  {author} {\bibinfo {author} {\bibfnamefont {P.}~\bibnamefont
  {Jain}}, \bibinfo {author} {\bibfnamefont {S.}~\bibnamefont {Panda}}, \ and\
  \bibinfo {author} {\bibfnamefont {S.}~\bibnamefont {Sarala}},\ }\href
  {\doibase 10.1103/PhysRevD.66.085007} {\bibfield  {journal} {\bibinfo
  {journal} {Phys. Rev. D}\ }\textbf {\bibinfo {volume} {66}},\ \bibinfo
  {pages} {085007} (\bibinfo {year} {2002})},\ \Eprint
  {http://arxiv.org/abs/hep-ph/0206046} {arXiv:hep-ph/0206046 [hep-ph]}
  \BibitemShut {NoStop}%
\bibitem [{\citenamefont {Agarwal}\ \emph {et~al.}(2008)\citenamefont
  {Agarwal}, \citenamefont {Jain}, \citenamefont {McKay},\ and\ \citenamefont
  {Ralston}}]{polarization:3}%
  \BibitemOpen
  \bibfield  {author} {\bibinfo {author} {\bibfnamefont {N.}~\bibnamefont
  {Agarwal}}, \bibinfo {author} {\bibfnamefont {P.}~\bibnamefont {Jain}},
  \bibinfo {author} {\bibfnamefont {D.~W.}\ \bibnamefont {McKay}}, \ and\
  \bibinfo {author} {\bibfnamefont {J.~P.}\ \bibnamefont {Ralston}},\ }\href
  {\doibase 10.1103/PhysRevD.78.085028} {\bibfield  {journal} {\bibinfo
  {journal} {Phys. Rev. D}\ }\textbf {\bibinfo {volume} {78}},\ \bibinfo
  {pages} {085028} (\bibinfo {year} {2008})},\ \Eprint
  {http://arxiv.org/abs/0807.4587} {arXiv:0807.4587 [hep-ph]} \BibitemShut
  {NoStop}%
\bibitem [{\citenamefont {Payez}\ \emph {et~al.}(2010)\citenamefont {Payez},
  \citenamefont {Cudell},\ and\ \citenamefont
  {Hutsem\'ekers}}]{polarization:5}%
  \BibitemOpen
  \bibfield  {author} {\bibinfo {author} {\bibfnamefont {A.}~\bibnamefont
  {Payez}}, \bibinfo {author} {\bibfnamefont {J.~R.}\ \bibnamefont {Cudell}}, \
  and\ \bibinfo {author} {\bibfnamefont {D.}~\bibnamefont {Hutsem\'ekers}},\
  }\href {\doibase 10.1063/1.3462669} {\bibfield  {journal} {\bibinfo
  {journal} {AIP Conference Proceedings}\ }\textbf {\bibinfo {volume} {1241}},\
  \bibinfo {pages} {444} (\bibinfo {year} {2010})},\ \Eprint
  {http://arxiv.org/abs/0911.3145} {arXiv:0911.3145 [astro-ph.CO]} \BibitemShut
  {NoStop}%
\bibitem [{\citenamefont {Agarwal}\ \emph {et~al.}(2011)\citenamefont
  {Agarwal}, \citenamefont {Kamal},\ and\ \citenamefont
  {Jain}}]{polarization:6}%
  \BibitemOpen
  \bibfield  {author} {\bibinfo {author} {\bibfnamefont {N.}~\bibnamefont
  {Agarwal}}, \bibinfo {author} {\bibfnamefont {A.}~\bibnamefont {Kamal}}, \
  and\ \bibinfo {author} {\bibfnamefont {P.}~\bibnamefont {Jain}},\ }\href
  {\doibase 10.1103/PhysRevD.83.065014} {\bibfield  {journal} {\bibinfo
  {journal} {Phys. Rev. D}\ }\textbf {\bibinfo {volume} {83}},\ \bibinfo
  {pages} {065014} (\bibinfo {year} {2011})},\ \Eprint
  {http://arxiv.org/abs/0911.0429} {arXiv:0911.0429 [hep-ph]} \BibitemShut
  {NoStop}%
\bibitem [{\citenamefont {Payez}\ \emph {et~al.}(2011)\citenamefont {Payez},
  \citenamefont {Cudell},\ and\ \citenamefont
  {Hutsem\'ekers}}]{polarization:7}%
  \BibitemOpen
  \bibfield  {author} {\bibinfo {author} {\bibfnamefont {A.}~\bibnamefont
  {Payez}}, \bibinfo {author} {\bibfnamefont {J.~R.}\ \bibnamefont {Cudell}}, \
  and\ \bibinfo {author} {\bibfnamefont {D.}~\bibnamefont {Hutsem\'ekers}},\
  }\href {\doibase 10.1103/PhysRevD.84.085029} {\bibfield  {journal} {\bibinfo
  {journal} {Phys. Rev. D}\ }\textbf {\bibinfo {volume} {84}},\ \bibinfo
  {pages} {085029} (\bibinfo {year} {2011})},\ \Eprint
  {http://arxiv.org/abs/1107.2013} {arXiv:1107.2013 [astro-ph.CO]} \BibitemShut
  {NoStop}%
\bibitem [{\citenamefont {Agarwal}\ \emph {et~al.}(2012)\citenamefont
  {Agarwal}, \citenamefont {Aluri}, \citenamefont {Jain} \emph
  {et~al.}}]{polarization:8}%
  \BibitemOpen
  \bibfield  {author} {\bibinfo {author} {\bibfnamefont {N.}~\bibnamefont
  {Agarwal}}, \bibinfo {author} {\bibfnamefont {P.~K.}\ \bibnamefont {Aluri}},
  \bibinfo {author} {\bibfnamefont {P.}~\bibnamefont {Jain}},  \emph {et~al.},\
  }\href {\doibase 10.1140/epjc/s10052-012-1928-y} {\bibfield  {journal}
  {\bibinfo  {journal} {The European Physical Journal C}\ }\textbf {\bibinfo
  {volume} {72}},\ \bibinfo {pages} {1928} (\bibinfo {year} {2012})},\ \Eprint
  {http://arxiv.org/abs/1108.3400} {arXiv:1108.3400 [astro-ph.CO]} \BibitemShut
  {NoStop}%
\bibitem [{\citenamefont {Payez}\ \emph {et~al.}(2012)\citenamefont {Payez},
  \citenamefont {Cudell},\ and\ \citenamefont
  {Hutsem\'ekers}}]{polarization:9}%
  \BibitemOpen
  \bibfield  {author} {\bibinfo {author} {\bibfnamefont {A.}~\bibnamefont
  {Payez}}, \bibinfo {author} {\bibfnamefont {J.}~\bibnamefont {Cudell}}, \
  and\ \bibinfo {author} {\bibfnamefont {D.}~\bibnamefont {Hutsem\'ekers}},\
  }\href {\doibase 10.1088/1475-7516/2012/07/041} {\bibfield  {journal}
  {\bibinfo  {journal} {J. Cosmol. Astropart. Phys}\ }\textbf {\bibinfo
  {volume} {2012}},\ \bibinfo {pages} {041} (\bibinfo {year} {2012})},\ \Eprint
  {http://arxiv.org/abs/1204.6187} {arXiv:1204.6187 [astro-ph.CO]} \BibitemShut
  {NoStop}%
\bibitem [{\citenamefont {Horns}\ \emph
  {et~al.}(2012{\natexlab{b}})\citenamefont {Horns}, \citenamefont {Maccione},
  \citenamefont {Mirizzi},\ and\ \citenamefont {Roncadelli}}]{polarization:10}%
  \BibitemOpen
  \bibfield  {author} {\bibinfo {author} {\bibfnamefont {D.}~\bibnamefont
  {Horns}}, \bibinfo {author} {\bibfnamefont {L.}~\bibnamefont {Maccione}},
  \bibinfo {author} {\bibfnamefont {A.}~\bibnamefont {Mirizzi}}, \ and\
  \bibinfo {author} {\bibfnamefont {M.}~\bibnamefont {Roncadelli}},\ }\href
  {\doibase 10.1103/PhysRevD.85.085021} {\bibfield  {journal} {\bibinfo
  {journal} {Phys. Rev. D}\ }\textbf {\bibinfo {volume} {85}},\ \bibinfo
  {pages} {085021} (\bibinfo {year} {2012}{\natexlab{b}})},\ \Eprint
  {http://arxiv.org/abs/1203.2184} {arXiv:1203.2184 [astro-ph.HE]} \BibitemShut
  {NoStop}%
\bibitem [{\citenamefont {Payez}(2013)}]{polarization:11}%
  \BibitemOpen
  \bibfield  {author} {\bibinfo {author} {\bibfnamefont {A.}~\bibnamefont
  {Payez}},\ }\href@noop {} {\  (\bibinfo {year} {2013})},\ \Eprint
  {http://arxiv.org/abs/1309.6114} {arXiv:1309.6114 [astro-ph.CO]} \BibitemShut
  {NoStop}%
\bibitem [{\citenamefont {Gong}\ \emph {et~al.}(2017)\citenamefont {Gong},
  \citenamefont {Chen} \emph {et~al.}}]{polarization:12}%
  \BibitemOpen
  \bibfield  {author} {\bibinfo {author} {\bibfnamefont {Y.}~\bibnamefont
  {Gong}}, \bibinfo {author} {\bibfnamefont {X.}~\bibnamefont {Chen}},  \emph
  {et~al.},\ }\href {\doibase 10.1103/PhysRevLett.118.061101} {\bibfield
  {journal} {\bibinfo  {journal} {Phys. Rev. Lett.}\ }\textbf {\bibinfo
  {volume} {118}},\ \bibinfo {pages} {061101} (\bibinfo {year} {2017})},\
  \Eprint {http://arxiv.org/abs/1612.05697} {arXiv:1612.05697 [astro-ph.HE]}
  \BibitemShut {NoStop}%
\bibitem [{\citenamefont {Brockway}\ \emph {et~al.}(1996)\citenamefont
  {Brockway}, \citenamefont {Carlson},\ and\ \citenamefont
  {Raffelt}}]{experiment1}%
  \BibitemOpen
  \bibfield  {author} {\bibinfo {author} {\bibfnamefont {J.~W.}\ \bibnamefont
  {Brockway}}, \bibinfo {author} {\bibfnamefont {E.~D.}\ \bibnamefont
  {Carlson}}, \ and\ \bibinfo {author} {\bibfnamefont {G.~G.}\ \bibnamefont
  {Raffelt}},\ }\href {\doibase 10.1016/0370-2693(96)00778-2} {\bibfield
  {journal} {\bibinfo  {journal} {Phys. Lett. B}\ }\textbf {\bibinfo {volume}
  {383}},\ \bibinfo {pages} {439 } (\bibinfo {year} {1996})},\ \Eprint
  {http://arxiv.org/abs/astro-ph/9605197} {arXiv:astro-ph/9605197 [astro-ph]}
  \BibitemShut {NoStop}%
\bibitem [{\citenamefont {Grifols}\ \emph {et~al.}(1996)\citenamefont
  {Grifols}, \citenamefont {Mass\'o},\ and\ \citenamefont
  {Toldr\`a}}]{experiment2}%
  \BibitemOpen
  \bibfield  {author} {\bibinfo {author} {\bibfnamefont {J.~A.}\ \bibnamefont
  {Grifols}}, \bibinfo {author} {\bibfnamefont {E.}~\bibnamefont {Mass\'o}}, \
  and\ \bibinfo {author} {\bibfnamefont {R.}~\bibnamefont {Toldr\`a}},\ }\href
  {\doibase 10.1103/PhysRevLett.77.2372} {\bibfield  {journal} {\bibinfo
  {journal} {Phys. Rev. Lett.}\ }\textbf {\bibinfo {volume} {77}},\ \bibinfo
  {pages} {2372} (\bibinfo {year} {1996})},\ \Eprint
  {http://arxiv.org/abs/astro-ph/9606028} {arXiv:astro-ph/9606028 [astro-ph]}
  \BibitemShut {NoStop}%
\bibitem [{\citenamefont {Payez}\ \emph {et~al.}(2015)\citenamefont {Payez},
  \citenamefont {Evoli}, \citenamefont {Fischer}, \citenamefont {Mirizzi},\
  and\ \citenamefont {Ringwald}}]{experiment5}%
  \BibitemOpen
  \bibfield  {author} {\bibinfo {author} {\bibfnamefont {A.}~\bibnamefont
  {Payez}}, \bibinfo {author} {\bibfnamefont {C.}~\bibnamefont {Evoli}},
  \bibinfo {author} {\bibfnamefont {M.}~\bibnamefont {Fischer}, \bibfnamefont
  {Tobias~Giannotti}}, \bibinfo {author} {\bibfnamefont {A.}~\bibnamefont
  {Mirizzi}}, \ and\ \bibinfo {author} {\bibfnamefont {A.}~\bibnamefont
  {Ringwald}},\ }\href {\doibase 10.1088/1475-7516/2015/02/006} {\bibfield
  {journal} {\bibinfo  {journal} {J.Cosmol.Astropart.Phys}\ }\textbf {\bibinfo
  {volume} {2015}},\ \bibinfo {pages} {006} (\bibinfo {year} {2015})},\ \Eprint
  {http://arxiv.org/abs/1410.3747} {arXiv:1410.3747 [astro-ph.HE]} \BibitemShut
  {NoStop}%
\end{thebibliography}%

\end{document}